\documentclass[11pt]{article}

\usepackage[fleqn]{amsmath}
\usepackage{color}
\usepackage{graphicx}
\usepackage{hyperref}
\usepackage{authblk}
\usepackage{times}
\usepackage{bm}                	
\usepackage{amssymb}         	
\usepackage{mathtools}
\DeclarePairedDelimiter{\norm}{\lVert}{\rVert}
\usepackage{braket}		
\usepackage{mathrsfs}                
\usepackage{amsthm}
\newtheorem{thm}{Theorem}

\newtheorem{lem}[thm]{Lemma}
\newtheorem{prop}[thm]{Proposition}
\newtheorem{exmp}[thm]{Example}
\newtheorem{defn}[thm]{Definition}
\usepackage{graphicx}
\usepackage{tikz}
\usetikzlibrary{calc, shapes, backgrounds,arrows,automata,backgrounds,fit,decorations.pathreplacing}
\usepackage[letterpaper,left=1in,top=1in,bottom=1in,right=1in,includeheadfoot,headheight=.6pt]{geometry}

\begin{document}

\title{Quantum games: a review of the history, current state, and interpretation}
\author{Faisal Shah Khan\footnote{Center on Cyber Physical Systems and Department of Mathematics, Khalifa University, Abu Dhabi, UAE.}, Neal Solmeyer \& Radhakrishnan Balu\footnote{Army Research Lab, Adelphi, MD, USA.}, 
Travis S.~ Humble\footnote{Quantum Computing Institute, Oak Ridge National Lab TN, USA. This manuscript has been authored by UT-Battelle, LLC under Contract No. DE-AC05-00OR22725 with the U.S. Department of Energy. The United States Government retains and the publisher, by accepting the article for publication, acknowledges that the United States Government retains a non-exclusive, paid-up, irrevocable, world-wide license to publish or reproduce the published form of this manuscript, or allow others to do so, for United States Government purposes. The Department of Energy will provide public access to these results of federally sponsored research in accordance with the DOE Public Access Plan (http://energy.gov/downloads/doe-public-access-plan).}}
\maketitle

\begin{abstract}
We review both theoretical and experimental developments in the area of quantum games since the inception of the subject circa 1999. We will also offer a narrative on the controversy that surrounded the subject in its early days, and how this controversy has affected the development of the subject.  
\end{abstract}

\section{Introduction}\label{intro}

Roots of the theory of quantum information lie in the ideas of Wisener and Ingarden \cite{Wiesner1983, Ingarden1976}. These authors proposed ways to incorporate the uncertainty or entropy of information within the uncertainty inherent in quantum mechanical processes. Uncertainty is a fundamental  concept in both physics and the theory of information, hence serving as the natural link between the two disciplines. Uncertainty itself is defined in terms of probability distributions. Every quantum physical object produces a probability distribution when its state is measured. More precisely, a general state of an $m$-state quantum object is an element of the $m$-dimensional complex projective Hilbert space $\mathbb{C}P^{m}$, say
\begin{equation}
v=\left(v_1, \dots, v_m \right).
\end{equation}
Upon measurement with respect to the observable states of the quantum object (which are the elements of an orthogonal basis of $\mathbb{C}P^{m}$), $v$ will produce the probability distribution 
\begin{equation}
\frac{1}{\sum_{k=1}^m|v_k|^2}\left(|v_1|^2, \dots, |v_m|^2 \right)
\end{equation}
over the observable states. Hence, a notion of quantum entropy or uncertainty may be defined that coincides with the corresponding notion in classical information theory \cite{nielsen2000quantum}. 

Quantum information theory was further developed as the theory of quantum computation by Feynman and Deutsch \cite{Feynman1982, Deutsch1985}. Feynman outlines a primitive version of what is now known as an $n$ qubit quantum computer \cite{nielsen2000quantum}, that is, a physical system that emulates any unitary function
\begin{equation}\label{unifunct}
\mathcal{Q}: \otimes_{i=1}^n\mathbb{C}P^{1}_i \longrightarrow \otimes_{i=1}^n\mathbb{C}P^{1}_i,
\end{equation}
where $\mathbb{C}P^{1}$ is the two-dimensional complex projective Hilbert space modeling the two-state quantum system or the qubit, in a way that $\mathcal{Q}$ can be expressed as a tensor product 
\begin{equation}
\mathcal{Q}= \otimes_{j=1}^n \mathcal{Q}_j
\end{equation}
of unitary functions $\mathcal{Q}_j$ (also known as quantum logic gates) that act only on one or two qubits \cite{PhysRevA.52.3457, PhysRevLett.94.230502}. Feynman's argument showed that it is possible to simulate two-state computations by Bosonic two-state quantum systems. 

Quantum computers crossed the engineering and commercialization thresholds in this decade with the Canadian technology company D-wave producing and selling a quantum annealing based quantum computer, and establish technology industry giants like IBM, Intel, Google, and Microsoft devoting financial resources to initiate their own quantum computing efforts. More generally, quantum information theory has made great strides starting in the 1980's in form of quantum communication protocols where, roughly speaking, one studies the transmission of quantum information over channels and applications. A milestone of quantum information theory is provably secure public key distribution which uses the uncertainty inherent in quantum mechanics to guarantee the security. This idea was first proposed by Charles Bennett and Gilles Brassard in 1984 at a conference in Bengaluru, India, and recently appeared in a journal \cite{Bennett2014}. Several companies including Toshiba and IDquantique offer commercial devices that can used to set up quantum cryptography protocols. While the literature on quantum information theory is vast, we refer the reader to books\cite{barnett2009,wilde2013} for further survey of quantum information theory. 

In the emerging field of quantum information technology, optimal implementation of quantum information processes will be of fundamental importance. To this end, the classic problem of optimizing a function's value will play a crucial role, with one looking to optimize the functional description of the quantum processes, as in  \cite{Miyake2001Geometric}, for example. Generalizing further, solutions to the problem of simultaneous optimization of two or more functions will be even more crucial given the uncertainty inherent in quantum systems. This generalized, multi-objective optimization problem forms the essence of non-cooperative game theory, where notional players are introduced as having the said functions as payoff or objective functions. The original single-objective optimization problem can also be studied as a single player game or what is also known as a dynamic game. 

We will give a mathematically formal discussion of non-cooperative games and their quantum mechanical counterparts in sections \ref{sec::noncoopgames} and \ref{sec::quantum games}, followed by a discussion in section \ref{sec::noncop} on the history of how such quantum games have been viewed and criticized in the literature in the context of the optimal implementation of quantum technologies as well the quantum mechanical implementation of non-cooperative games. In section \ref{coop} we contrast cooperative and non-cooperative games, and in section \ref{social}, we introduce a new perspective on quantum entanglement as a mechanism for establishing social equilibrium. Section \ref{quantumalgo} gives an overview of quantum games and quantum algorithms and communication protocols, and section \ref{Bell} concerns Bell's inequalities and their role in quantum Bayesian games, and sections \ref{Stochastic} through \ref{QSG} concern classical and quantum versions of stochastic and dynamic games. We give the current state-of-affairs in the experimental realization of quantum games in section \ref{exp}, followed by section \ref{future} that discusses potential future applications of quantum games.

\section{Non-cooperative games}\label{sec::noncoopgames}

Non-cooperative game theory is the mathematical foundation of making optimal decisions in competitive situations based on available information. The written philosophical foundations of Game Theory trace back to at least the great works of Sun Tsu ({\it The Art of War}), circa 500 BCE in China, and Chanakya ({\it Arthashastra}), circa 250 BCE in India. Sun Tsu captures the essence of game-theoretic thinking in the following (translated \cite{tzu2002art}) lines from {\it The Art of War}: 

\begin{quote}
Knowing the other and knowing oneself,
	In one hundred battle no danger, \\
Not knowing the other and knowing oneself, 
One victory for one loss, \\
Not knowing the other and not knowing oneself,
In every battle certain defeat (Denma translation).
\end{quote}

In short, each competitor or {\it player}, in the competitive situation or {\it game}, should know the preferences of each player over the outcomes of the game, and knowing this information is sufficient for each player to make optimal decisions or {\it strategic choices}. The word ``optimal'' requires further elaboration. In non-cooperative game theory, there are two ways to utilize it.  

First is via the notion of Nash equilibrium, proposed by Nobel Laureate John Nash \cite{Nash1950}, where each player's strategic choice, {\it given} the strategic choices of all the other players, produces an outcome of the game that maximizes the player's preferences over the outcomes. In other words, unilateral deviation by the player to another strategic choice will produce an outcome which is less preferable to the player. Further yet, one can say that each player's strategic choice is a {\it best response} to every other. The second way the word ``optimal'' is used in game theory is via the notion of Pareto-optimality where the strategic choices made by the players produce an outcome of the game that maximizes the preferences of {\it every} player. In other words, a unilateral deviation by any one player to some other strategic choice will produce an outcome which is less preferred by {\it some} player. If the adversely affected player is also the one who unilaterally deviated, then the Pareto-optimal outcome is also a Nash equilibrium. Note that Nash equilibrium is a more likely outcome in a non-cooperative game than a Pareto-optimal one in the sense that on average, or in repeated games, players' strategy choices will tend toward the Nash equilibrium.

 Formalizing, we say that an $N$ player, non-cooperative game in normal form is a function $\Gamma$ 
\begin{equation}\label{Game}
\Gamma: \prod_{i=1}^N S_i \longrightarrow O,
\end{equation}
with the additional feature of the notion of non-identical preferences over the elements of the set of {\it outcomes} $O$, for every ``player'' of the game. The preferences are a pre-ordering of the elements of $O$, that is, for $l,m,n \in O$
\begin{equation}
m \preceq m,  \hspace{2mm}  {\rm and}  \hspace{2mm}  l \preceq m \hspace{2mm} {\rm and} \hspace{2mm}  m \preceq n \implies  l \preceq n.
\end{equation} 
where the symbol $\preceq$ denotes ``of less or equal preference''. Preferences are typically quantified numerically for the ease of calculation of the payoffs. To this end, functions $\Gamma_i$ are introduced which act as the {\it payoff function} for each player $i$ and typically map elements of $O$ into the real numbers in a way that preserves the preferences of the players. That is, $\preceq$ is replaced with $\leq$ when analyzing the payoffs. The factor $S_i$ in the domain of $\Gamma$  is said to be the {\it strategy set} of player $i$, and a {\it play} of $\Gamma$ is an $n$-tuple of strategies, one per player, producing a payoff to each player in terms of his preferences over the elements of $O$ in the image of $\Gamma$.

A {\it Nash equilibrium} is a play of $\Gamma$ in which every player employs a strategy that is a {\it best reply}, with respects to his preferences over the outcomes, to the strategic choice of every other player. In other words, unilateral deviation from a Nash equilibrium by any one player in the form of a different choice of strategy will produce an outcome which is less preferred by that player than before. Following Nash, we say that a play $p'$  of $\Gamma$ {\it counters} another play $p$ if  $\Gamma_i(p') \geq \Gamma_i(p)$ for all players $i$, and that a self-countering play is an (Nash) equilibrium. 

Let $C_{p}$ denote the set of all the plays of $\Gamma$ that counter $p$. Denote $\prod_{i=1}^n S_i$ by $S$ for notational convenience, and note that $C_{p} \subset S$ and therefore $C_{p} \in 2^S$. Further note that the game $\Gamma$ can be factored as  
\begin{equation}\label{factor}
\Gamma:  S \xrightarrow{\Gamma_C} 2^S \xrightarrow{E} O
\end{equation}
where to any play $p$ the map $\Gamma_C$ associates its countering set $C_p$ via the payoff functions $\Gamma_i$. The set-valued map $\Gamma_C$ may be viewed as a pre-processing stage where players seek out a self-countering play, and if one is found, it is mapped to its corresponding outcome in $O$ by the function $E$. The condition for the existence of a self-countering play, and therefore of a Nash equilibrium, is that $\Gamma_C$ have a fixed point, that is, an element $p^* \in S$ such that $p^* \in C_{p^*}$.

In a general set-theoretic setting for non-cooperative games, the map $\Gamma_C$ may not have a fixed point. Hence, not all non-cooperative games will have a Nash equilibrium. However, according to Nash's theorem, when the $S_i$ are finite and the game is extended to its {\it mixed} version, that is, the version in which randomization via probability distributions is allowed over the elements of all the $S_i$, as well as over the elements of $O$, then $\Gamma_C$ has at least one fixed point and therefore at least one Nash equilibrium. 

Formally, given a game $\Gamma$ with finite $S_i$ for all $i$, its mixed version is the product function 
\begin{equation}\label{mixedgame} 
\Lambda: \prod_{i=1}^n \Delta(S_i) \longrightarrow \Delta(O)
\end{equation}
where $\Delta(S_i)$ is the set of probability distributions over the $i^{\rm{th}}$ player's strategy set $S_i$, and the set $\Delta(O)$ is the set of probability distributions over the outcomes $O$. Payoffs are now calculated as expected payoffs, that is, weighted averages of the values of $\Gamma_i$, for each player $i$, with respect to probability distributions in $\Delta(O)$ that arise as the product of the plays of $\Lambda$. Denote the expected payoff to player $i$ by the function $\Lambda_i$. Also, note that $\Lambda$ restricts to $\Gamma$.
In such $n$-player games, at least one Nash equilibrium play is guaranteed to exist as a fixed point of $\Lambda$ via Kakutani's fixed-point theorem \cite{Kakutani1941}.
\\

\noindent {\bf Kakutani's fixed-point theorem}: {\it Let $S \subset \mathbb{R}^n$ be nonempty, bounded, closed, and convex, and let $F: S\rightarrow 2^S$ be an upper semi-continuous set-valued mapping such that $F(s)$ is non-empty, closed, and convex for all $s \in S$. Then there exists some $s^* \in S$ such that $s^* \in F(s^*)$}.
\\

To see this, make $S=\prod_{i=1}^n \Delta(S_i)$. Then $S \subset \mathbb{R}^n$ and $S$ is non-empty, bounded, and closed because it is a finite product of finite non-empty sets. The set $S$ is also convex because its the convex hull of the elements of a finite set. Next, let $C_p$ be the set of all plays of $\Lambda$ that counter the play $p$. Then $C_p$ is non-empty, closed, and convex. Further, $C_p \subset S$ and therefore $C_p \in 2^S$. Since $\Lambda$ is a game, it factors according to (\ref{factor})
\begin{equation}
\Lambda: S \xrightarrow{\Lambda_C} 2^S \xrightarrow{E_{\Pi}} \Delta(O)
\end{equation}
where the map $\Lambda_C$ associates a play to its countering set via the payoff functions $\Lambda_i$. Since $\Lambda_i$ are all continuous, $\Lambda_C$ is continuous. Further, $\Lambda_C(s)$ is non-empty, closed, and convex for all $s \in S$ (we will establish the convexity of $\Lambda_C(s)$ below; the remaining conditions are also straightforward to establish). Hence, Kakutani's theorem applies and there exists an $s^* \in S$ that counters itself, that is, $s^* \in \Lambda_C(s^*)$, and is therefore a Nash equilibrium. The function $E_{\Pi}$ simply maps $s^*$ to $\Delta(O)$ as the product probability distribution from which the Nash equilibrium expected payoff is computed for each player. 

The convexity of the $\Lambda_C(s)=C_p$ is straight forward to show. Let $q,r \in C_p$. Then
\begin{equation}\label{counteringeq}
\Lambda_i(q) \geq \Lambda_i(p) \quad {\rm and} \quad \Lambda_i(r) \geq \Lambda_i(p)
\end{equation} 
for all $i$.
Now let $0 \leq \mu \leq 1$ and consider the convex combination $\mu q + (1-\mu)r$ which we will show to be in $C_p$. First note that $\mu q + (1-\mu)r \in S$ because $S$ is the product of the convex sets $\Delta(S_i)$. Next, since the $\Lambda_i$ are all linear, and because of the inequalities in (\ref{counteringeq}) and the restrictions on the values of $\mu$,
\begin{equation}\label{linearity}
\Lambda_i(\mu q+ (1- \mu)r)=\mu \Lambda_i(q) + (1- \mu) \Lambda_i(r) \geq \Lambda_i(p)
\end{equation}
whereby $\mu q + (1-\mu)r \in C_p$ and $C_p$ is convex. Going back to the game $\Gamma$ in (\ref{Game}) defined in the general set-theoretic setting, certainly Kakutani's theorem would apply to $\Gamma$ if the conditions are right, such as when the image set of $\Gamma$ is pre-ordered and $\Gamma$ is linear.

Kakutani's fixed-point theorem can be generalized to include subsets $S$ of convex topological vector spaces, as was done by Glicksberg in \cite{Glicksberg1952}. The following is a paraphrased but equivalent statement of Glicksberg's fixed-point theorem (the term ``linear space'' in the original statement of Glicksberg's theorem is equivalent to the term vector space):
\\

\noindent {\bf Glicksberg's fixed-point theorem}: {\it Let $H$ be nonempty, compact, convex subset of a convex Hausdorff topological vector space and let $\Phi: H\rightarrow 2^H$ be an upper semi-continuous set-valued mapping such that $\Phi(h)$ is non-empty and convex for all $h \in H$. Then there exists some $h^* \in H$ such that $h^* \in \Phi(h^*)$}.
\\

\noindent Using Glicksberg's fixed-point theorem, one can show that Nash equilibrium exists in games where the strategy sets are infinite or possibly also un-countably infinite.

Non-cooperative game theory has been an immensely successful mathematical model for studying scientific and social phenomena. In particular, it has offered key insights into equilibrium and optimal behavior in economics, evolutionary biology, and politics. As with any established subject, game theory has a vast literature available. However, we refer the reader to \cite{binmore2007playing,myerson1991game}. Given the successful interface of non-cooperative game theory with several other subjects, it is no wonder then that physicist have explored the possibility of using Game Theory to model physical processes as games and study their equilibrium behaviors. The first paper that the author's are aware of in which aspects of quantum physics, wave mechanics in particular, were viewed as games is \cite{Blaquiere1980}. 

\section{Non-cooperative quantum games}\label{sec::quantum games}

\begin{figure}
\centering
\includegraphics{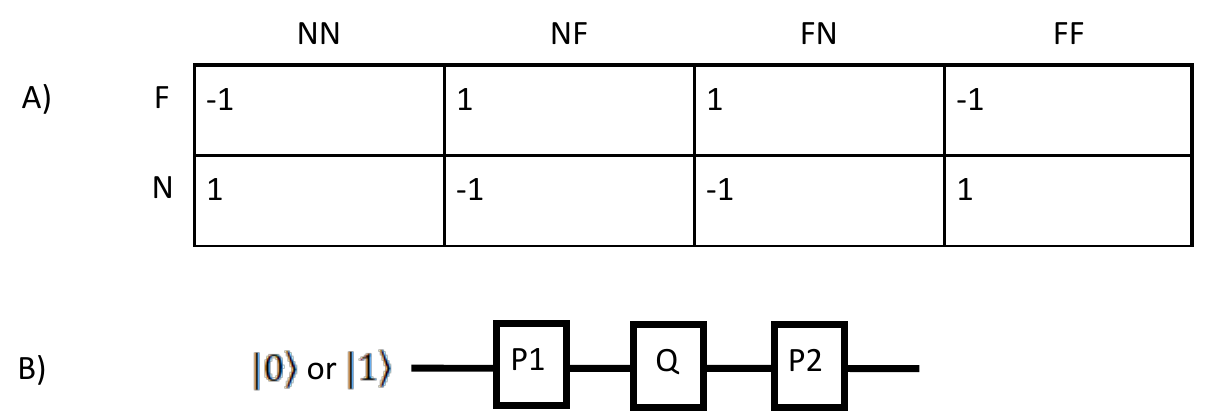}
\caption{{\small A) The Penny Flip game in normal form, with N and F representing the players' strategic choices of Flip and No-Flip respectively. B) The quantum circuit for this game, with P1 and P2 being quantum strategies of Player 1, and Q being the quantum strategy of Player 2, which when chosen to be the Hadamard gate $H$, allows Player 2 to always win the game when the input qubit state is either $\ket{0}$ or $\ket{1}$. } }
\label{PQ}
\end{figure}

A formal merger of non-cooperative game theory and quantum computing was initiated in \cite{PhysRevLett.82.1052} by D. Meyer, who was motivated to study efficient quantum algorithms, and to this end, proposed a game-theoretic model for quantum algorithms. To this end, his focus of study was the situation where in a particular two player game, one of the players had access to {\it quantum strategies}. Meyer in fact did not introduce the term ``quantum game'' in his work; rather, this was done by another group of authors whose work will be discussed shortly. Meyer defined a quantum strategy to be a single qubit logic gate in the quantum computation for which the game model was constructed. The particular game he considers is the Penny Flip game of Figure \ref{PQ}A), which he then realizes as a single qubit quantum circuit of Figure \ref{PQ}B) in which the first player employs P1 and P2 from the restricted set of quantum operations
$$
\left\{  \left[ {\begin{array}{cc}
   1 & 0 \\
   0 & 1 \\
  \end{array} } \right],  \left[ {\begin{array}{cc}
   0 & 1 \\
  1 &  0 \\
  \end{array} } \right] \right\}
$$
whereas Player 2 is allowed to employ, in particular, the Hadamard operation
$$
H= \frac{1}{\sqrt{2}}\left[ {\begin{array}{cc}
   1 & -1 \\
   -1 & -1 \\
  \end{array} } \right],
$$
both on either the qubit state $\ket{0}$ or $\ket{1}$. When the quantum circuit is played out with respect to the computational basis for the Hilbert space, one sees that Player 2 always wins the game. 

A similar two player game model was applied to quantum algorithms such as Simon's and Grover's algorithms. Meyer showed that in this setting, the player with access to a proper quantum strategy (and not simply a classical one residing inside a quantum one) would always win this game. He further showed that if both players had access to proper quantum strategies, then in a strictly competitive or zero-sum game (where the preferences of the players over the outcomes are diametrically opposite), a Nash equilibrium need not exist. However, in the case where players are allowed to choose their quantum strategies with respect to a probability distribution, that is, employ mixed quantum strategies, Meyer used Glicksberg's fixed point theorem to show that in this situation Nash equilibrium would always exist. Meyer's work also provides a way to study equilibrium behavior of quantum computational mechanisms. 

\begin{figure}
\centering
\includegraphics{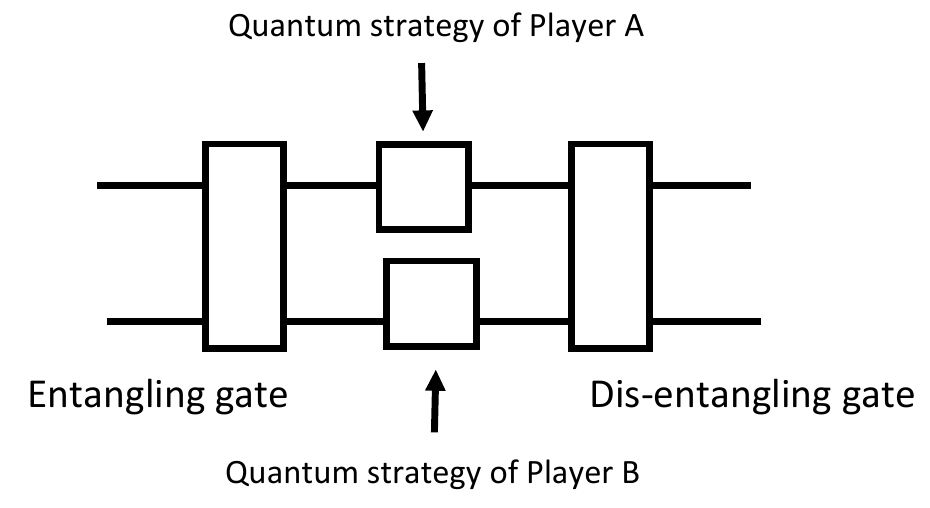}
\caption{{\small Quantum circuit for the EWL quantization scheme. The two qubit gate to the left of the players' quantum strategies maximally entangles the two qubits employed to play the game. The two qubit gate appearing after the quantum strategies dis-entangles the qubits with respect to the fixed, computational basis of the Hilbert space.} }
\label{EWL}
\end{figure}

\begin{figure}
\centering
\includegraphics{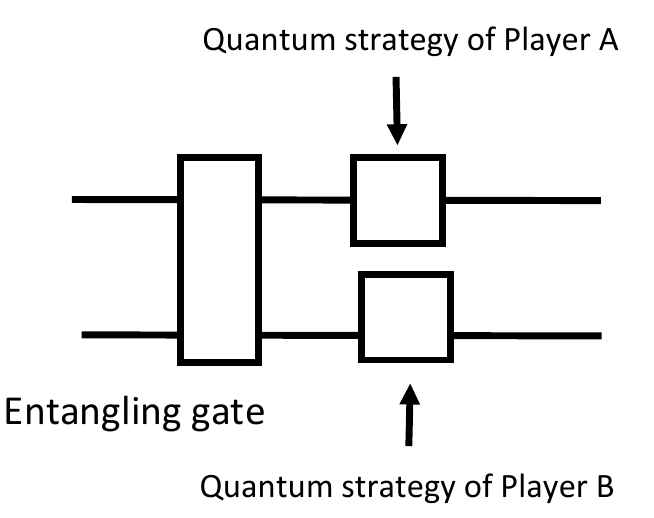}
\caption{{\small Quantum circuit for the MW quantization scheme. Although similar to the EWL scheme, this scheme does not restrict to the original classical game due to the lack of the dis-entangling operation before quantum measurement. This is due to the fact that the classical game is encoded into the Hilbert space as the elements of the computational basis.} }
\label{MW}
\end{figure}

The term {\it quantum game} appears to have been first used by Eisert, Wilkens, and Lewenstein in their paper \cite{PhysRevLett.83.3077} which was published soon after Meyer's work. These authors were interested in, as they put it, ``... the quantization of non-zero sum games''. At face value, this expression can create controversy (and it has), since quantization is a physical process whereas a game is primarily an abstract concept. However, Chess, Poker, and Football are examples of games that can be implemented {\it physically}. It becomes clear upon reading the paper that the authors' goal is to investigate the consequences of a non-cooperative game implemented {\it quantum physically}. More accurately, Eisert {\it et al.} give a quantum computational implementation of Prisoner's Dilemma. This implementation is reproduced in Figure \ref{EWL}.
Eisert {\it et al.} show that in their quantum computational implementation of the non-strictly competitive game of Prisoner's Dilemma, when followed by quantum measurement, players can achieve a Nash equilibrium that is also Pareto-optimal. One should view the ``EWL quantization protocol'' for Prisoner's Dilemma as an extension of the original game to include higher order randomization via quantum superposition and entanglement followed by measurement \cite{Bleiler2009} similar to the way game theorists have traditionally extended (or physically implemented) a game to include randomization via probability distributions. And indeed, Eisert {\it et al.} ensure that their quantum game restricts to the original Prisoner's Dilemma.  Inspired by the EWL quantization protocol, Marinatto and Weber proposed the ``MW quantization protocol'' in \cite{MARINATTO2000291}. As Figure \ref{MW} shows, the MW protocol differs from the EWL protocol in only the absence of the dis-entangling gate.
 
\begin{figure}
\centering
\includegraphics{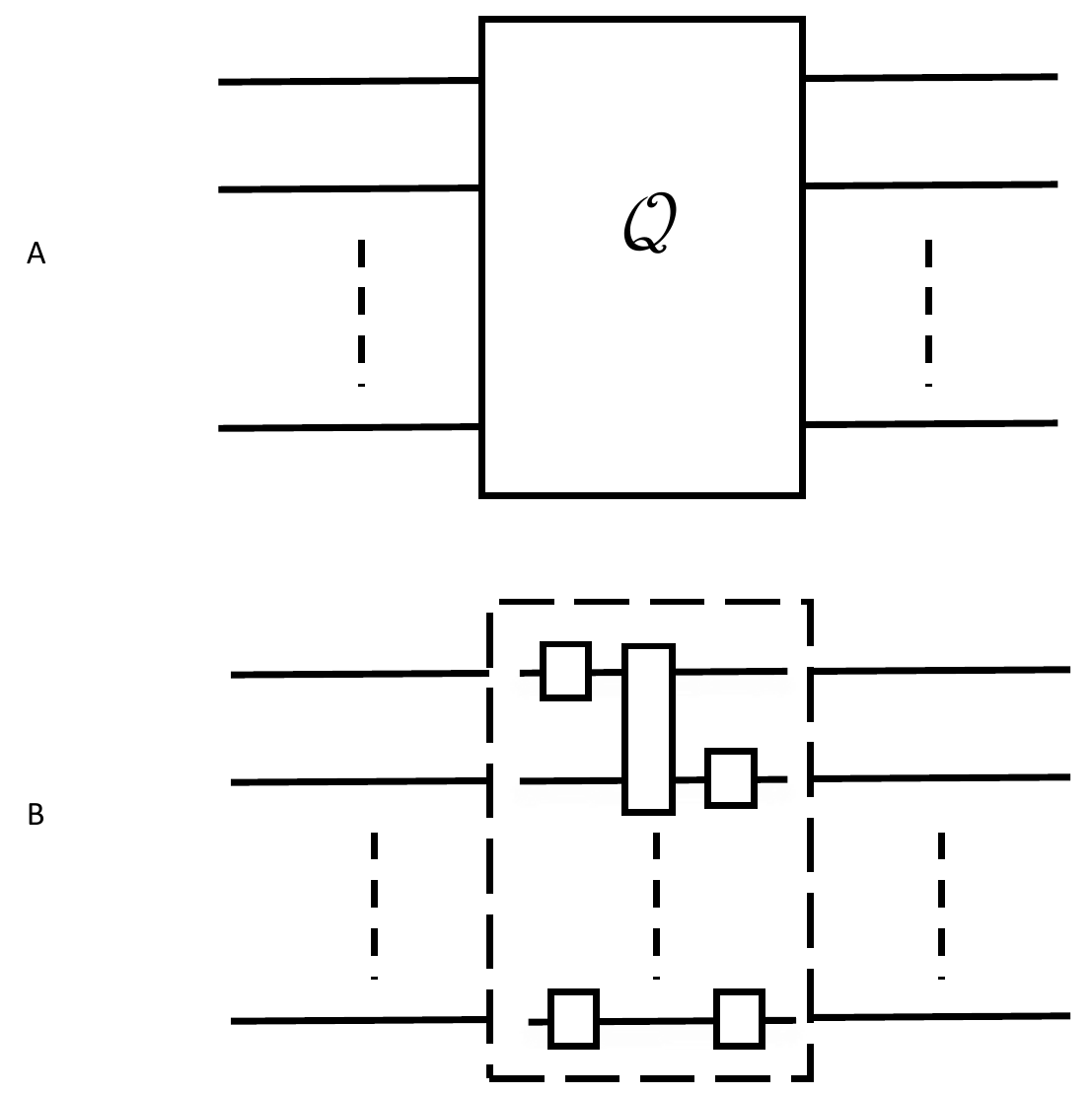}
\caption{{\small A) A $n$-player quantum game $\mathcal{Q}$ as a $n$ qubit quantum logic gate, with the provision that preferences are defined over the elements of the computational basis of the Hilbert space, one per player. B) An example of playing the quantum game $\mathcal{Q}$ (equivalently, implementing the quantum logic gate) as a quantum circuit comprised of only one qubit logic gates (strategies) and two qubit logic gates (quantum mediated communication) using matrix decomposition techniques such as the cosine-sine decomposition \cite{Sutton}.} }
\label{Qgame}
\end{figure}

Whereas Meyer's seminal work laid down the mathematical foundation of quantum games via its Nash equilibrium result using a fixed-point theorem, the works of Eisert {\it et al.} and Marinatto {\it et al.} have been the dominant protocols for quantization of games. But before discussing the impact of these works on the subject of quantum game theory, it is pertinent to introduce a mathematically formal definition of a non-cooperative quantum game in normal form that is consistent with these authors' perspectives. 
An $n$-player quantum game in normal form arises from (\ref{Game}) when one introduces quantum physically relevant restrictions. We define a pure strategy quantum game (in normal form) to be any unitary function
\begin{equation}\label{qgamedef}
\mathcal{Q}: \otimes_{i=1}^n \mathbb{C}P^{d_i} \longrightarrow \otimes_{i=1}^n  \mathbb{C}P^{d_i}
\end{equation}
where $\mathbb{C}P^{d_i}$ is the $d_i$-dimensional complex projective Hilbert space of pure quantum states that constitutes player $i$'s pure quantum strategies, as well as the set of outcomes with a notion of non-identical preferences defined over its elements, one per player \cite{FSKHAN20131}. Figure \ref{Qgame} captures this definition as a quantum circuit diagram. A mixed quantum game would then be any function
\begin{equation}\label{mixqgamedef}
\mathcal{R}: \Delta \left( \otimes_{i=1}^n \mathbb{C}P^{d_i} \right) \longrightarrow \Delta \left( \otimes_{i=1}^n  \mathbb{C}P^{d_i} \right)
\end{equation}
where $\Delta$ represents the set of probability distributions over the argument. Our definition of a quantum game in both (\ref{qgamedef}) and (\ref{mixqgamedef}) is consistent with Meyer's perspective in the sense that it allows one to constrained optimize a quantum mechanism by defining payoff functions before measurement, and it is consistent with the EWL perspective if one a defines the payoff functions after measurement as expected value.  
As mentioned earlier, Meyer used Glicksberrg's fixed point theorem to establish the guarantee of Nash equilibrium in the mixed quantum game $\mathcal{R}$. This is not surprising given that probabilistic mixtures form a convex structure, which is an essential ingredient for fixed-point theorems to hold on ``flat'' manifolds such as $\mathbb{R}^n$. However, it was only very recently that two of the current authors showed that Nash equilibrium via fixed-point theorem can be guaranteed in the quantum game $\mathcal{Q}$ \cite{Khan2018}. These authors used the Riemannian manifold structure of $\mathbb{C}P^{n}$ to invoke John Nash's other, mathematically more popular theorem known as the Nash embedding theorem:
\\

\noindent {\bf Nash embedding theorem}: {\it Every compact Riemannian manifold can be (isometrically) embedded into $\mathbb{R}^m$ for a sufficiently large $m$}. 
\\

\noindent The Nash embedding theorem tells us that $\mathbb{C}P^{n}$ is diffeomorphic to its image under a length preserving map into $\mathbb{R}^m$. With suitable considerations in place, it follows that Kakutani's theorem applies to the image of $\mathbb{C}P^{n}$ in $\mathbb{R}^m$. Now, tracing the diffeomorphim back to $\mathbb{C}P^{n}$ guarantees the existence of Nash equilibrium in the pure quantum game $\mathcal{Q}$. 

Another key insight established in \cite{Khan2018} is that just as in classical games, linearity of the payoff functions is a fundamental requirement for guaranteeing the existence of Nash equilibrium in pure quantum games. Hence, quantization of games such as the EWL protocol, in which the payoff is the expected value computed after quantum measurement, cannot guarantee Nash equilibrium. On the other hand, the problem of pure state preparation, when viewed as a quantum game with the overlap (measured by the inner-product) of quantum states as the payoff function, guarantees Nash equilibrium.

\section{Criticism of quantum games - a discussion}\label{sec::noncop}

Criticism of quantum games has historically been focused on the EWL quantization protocol and we will discuss this criticism in detail below. On the other hand, we conjecture that the reason that Meyer's work has not seen much criticism is its mathematically formal foundation, and we also note that the MW quantization protocol has not been subjected to the same level of scrutiny as the EWL protocol vis-a-vis quantum physical implementation of games. This is remarkable because the MW protocol {\it does not} restrict to the original classical two player game (Prisoners' Dilemma, for example) coded into the Hilbert space via identification with the elements of a fixed basis. Therefore, the MW protocol holds little game-theoretic meaning! Frackiewicz has produced an adaptation of the MW protocol in \cite{Frackiewicz} which attempts to rectify this protocol's deficiencies. Nonetheless, the original, game-theoretically questionable version of MW still appears in quantum game theory papers, for example \cite{Deng, SAMADI201894}. 

Although the MW quantization protocol lacks game-theoretic substance in its original form, it is amenable to interpretation as an example of applying non-cooperative game theory to quantum mechanics, or ``gaming the quantum'' \cite{FSKHAN2013}. In this interpretation, the MW protocol represents a game-theoretic approach to designing quantum computational mechanisms which exhibit optimal behavior under multiple constraints. This makes the MW protocol more akin to Meyer's approach of using game theory to gain insights into quantum algorithms for quantum computation and communication. We will discuss this idea further in section \ref{quantumalgo}. The remainder of this section is devoted to a discussion of the EWL quantization protocol and its criticism.

In \cite{PhysRevA.66.024306}, van Enk {\it et al.} state that the output of the EWL protocol for a specific and finite set of quantum strategies, after measurement, produces a function that is an extension of Prisoner's Dilemma but is entirely non-quantum mechanical. These authors argue that since this post measurement function emulates the results of the EWL quantization protocol, the quantum nature of the latter is redundant. However, if this criticism is taken seriously, then extensions of Prisoner's Dilemma via probability distributions can also be restricted to specific, finite mixed strategy sets to produce a larger game that is entirely non-probabilistic and which has a different structure than the original game! 

The source of this criticism appears to be a confusion between descriptive and prescriptive interpretations of game theory. For the mixed game should not be understood as a description of a game that utilizes piece-wise larger, non-probabilistic games. Rather, the reasoning behind extending to a mixed game is prescriptive, allowing one to design a mechanism that identifies probability distributions over the players' mixed strategies, which when mapped to probability distributions over the outcomes of the game via the product function, produce an expected outcome of the game as Nash equilibrium. From this point of view, the EWL protocol is a perfectly valid higher order mechanism for extending Prisoner's Dilemma.

Another criticism by van Enk {\it et al.} of the EWL quantization protocol is that it does not preserve the non-cooperative nature of Prisoner's Dilemma due to the presence of quantum entanglement generated correlations. Eisert {\it et al.} have argued that entanglement can be viewed as an honest referee who communicates to the players on their behalf. But van Enck {\it et al.} insist that introducing communication between the players ``...blurs the contrast between cooperative and non-cooperative games''. This is true, but classical game theory also has a long and successful history of blurring this distinction through the use of mediated communication. Bringing in an honest referee into a game is just another form of game extension known as {\it mediated communication} \cite{AUMANN197467} which, to be fair, can easily be mistaken as a form of cooperation. In fact however, such games are non-cooperative and Nash equilibrium still holds as the suitable solution concept. It is only when one tries to relate the Nash equilibrium in the extended game (with mediated communication) to a notion of equilibrium in the original game that the broader notion of {\it correlated equilibrium} arises. The motivation for introducing mediated communication in games comes from the desire to realize probability distributions over the outcomes of a game which are not in the image of the mixed game. From this  perspective, the EWL protocol could be interpreted as a higher order extension of Prisoner's Dilemma to include quantum mediated communication. An excellent, mathematically formal explanation of the latter interpretation can be found in \cite{Bleiler2009}. 

Finally, in \cite{PhysRevLett.87.069801}, Benjamin {\it et al.} argue that Nash equilibrium in the EWL protocol, while game-theoretically correct, is of limited quantum physical interest. These authors proceed to show that when a naturally more general and quantum physically more meaningful implementation of EWL is considered, the quantum Prisoner's Dilemma has no Nash equilibrium! However, once randomization via probability distributions is introduced into their quantum Prisoner's Dilemma, a Nash equilibrium that is near-optimal, but still better paying than the one available in the classical game, materializes again in line with Glicksberg-Meyer theorem. This criticism was in fact addressed by Eisert and Wilkens in a follow up publication \cite{doi:10.1080/09500340008232180}. Benjamin {\it et al.} give a discrete set of strategies that could be employed in a classical set-up of the game that gives the same solution to the dilemma. Eisert {\it et al}'s strategy set is then just the continuous analogue of this discrete set. 

One may contextualize the criticism of quantum computational implementation of games using the more formal language of computational complexity theory as follows. The class BQP is that of problems that can be efficiently solved on a quantum computer, and P is the class that can be solved efficiently on a classical computer. It is known that P is strictly contained in BQP and, moreover, that BQP is {\it not} contained in P. That is to say, there exist problems that quantum computers can solve efficiently but which classical computers cannot. Examples include quantum simulation, quantum search (in the Oracle setting), and solving systems of linear equations. However, it is currently unknown how BQP compares to BPP, the class of problems that can be solved on a probabilistic classical computer. The latter ambiguity calls into question whether efficient classical methods may exists for some quantum algorithms, such as the famous Shor's factoring algorithm. 

In particular, criticism of the EWL protocol may now be succinctly phrased as follows: it has been shown previously by Eisert {\it et al}. that some quantum games, such as Prisoner’s Dilemma, may provide better paying Nash equilibria when entanglement alone is added to the player’s strategies. However, van Enk and Pike have noted that permitting classical advice within those same games can recover similar Nash equilibria. This raises the question as to whether the power of quantum computational protocols for games using entanglement are captured completely by classical sprotocols using advice. This open question is akin to the question is quantum computing as to whether BQP is contained in BPP. 

We end this section by noting that the EWL protocol is just one of many possible quantization protocols that may be constructed via quantum circuit synthesis methods such as the one envisioned in Figure \ref{Qgame}. Consequences of these other quantum computational implementations of non-cooperative games used in economics, evolutionary biology, and any other subject where game theory is applicable, appear to be largely unexplored. 

\subsection{Quantum games: cooperative versus non-cooperative}\label{coop}

In the almost two decades since the inception of the theory of quantum games, the EWL quantization protocol has taken on the role of a working definition for non-cooperative quantum games for physicists. Several notable results on the quantum physical implementation of games followed Eiset {\it et al.}'s work, such as \cite{MARINATTO2000291,PhysRevA.64.030301,PhysRevA.63.020302, iqbal2001evolutionarily, PhysRevA.65.062318, PhysRevA.65.022306, DU2002229, 0305-4470-38-2-011, 0305-4470-37-22-012, PhysRevA.65.052320}. This may seem odd as one would think that the physics community would be more interested in the equilibrium or optimal behavior of quantum systems than the quantum physical implementation of games. On the other hand, this scenario makes sense from a practical point of view because with the advent of technological realizations of quantum computers and quantum communication systems, the playability of games quantum computationally would be of fundamental importance for financial and economic decision making. 

There is a considerable body of work in which the authors claim to {\it cooperatively} game the quantum. Several authors in the early 2000's, such as \cite{PhysRevA.67.012304, cleve2004consequences, PhysRevLett.87.217901, kkmtv08, Aharonov:2000:QBE:335305.335404, Marriott_quantumarthur-merlin}, gamed quantum communication protocols or studied complexity classes for quantum processes by considering the protocols as {\it cooperative} games of incomplete information. While most authors of such work have mainly focused on identifying quantum advantages similar to the one Meyer identified in his paper, the source of motivation for their work is different. For example, in \cite{PhysRevA.67.012304}, the authors state: ``Formally, a quantum coin flipping protocol with bias $\epsilon$ is a two-party communication game in the style of [15]...'' where the citation provided is to Chi-Chi Yao's paper titled {\it Quantum circuit complexity} \cite{Chi-ChihYao:1993:QCC:1398517.1398916}. Despite the fact that cooperative game theory is the motivation for the latter and other similar work, a formal discussion of cooperative games together with a formal mapping of the relevant physics to the requisite cooperative game is almost always missing. In fact, it would be accurate to say the the word ``game'' is thrown around in this body of literature as frivolously as the headless carcass of a goat is thrown around in the Afghan game of Buzkashi; but the beef is nowhere to be found. This is not surprising since beef comes from cows! The point of this somewhat macabre analogy is that one should be just as disturbed when hearing the word ``game'' used for an object that isn't one, as one surely is when hearing the word ``beef'' used for a goat carcass. 

Cooperative games are sophisticated conceptual and mathematical constructs. Quoting Aumann \cite{Aumann1989} (the quote is taken from \cite{Brandenburger})

\begin{quote}
Cooperative (game) theory starts with a formalization of games that
abstracts away altogether from procedures and $\dots$ concentrates, instead, on the possibilities for agreement $\dots$ There are several reasons that explain why cooperative games came to be treated separately.  One is that when one does build negotiation and enforcement procedures explicitly into the model, then the results of a non-cooperative analysis depend very strongly on the precise form of
the procedures, on the order of making offers and counter-offers and so on. This may be appropriate in voting situations in which precise rules of parliamentary order prevail, where a good strategist can indeed carry the day. But problems of negotiation are usually more amorphous; it is difficult to pin down just what the procedures are. More fundamentally, there is a feeling that procedures are not really all that relevant; that it is the possibilities for coalition forming, promising and threatening that are decisive, rather than whose turn it is to speak. $\dots$ Detail distracts attention from essentials. Some things are seen better from a distance; the Roman camps around Metzada are indiscernible when one is in them, but easily visible from the top of the mountain.
\end{quote}

More formally, a cooperative game allows players to benefit by forming coalitions, and binding agreements are possible. This means that a formal definition of a cooperative game is different than that of a non-cooperative one, and instead of Nash equilibrium, cooperative games entertain solution concepts such as a {\it coalition structure} consisting of various coalitions of players, together with a payoff vector for the various coalitions. Optimality features of the solution concepts are different than those in non-cooperative games as well. For instance, there is the notion of an {\it imputation} which is a coalition structure in which every player in a coalition prefers to stay in the coalition over ``going at it alone". While aspects of cooperative games are certainly reminiscent of those of non-cooperative games, the two types of games are very different objects in very different categories. Because the body of literature that purports to utilize cooperative games to identify some form of efficient or optimal solutions in quantum information processes does so via unclear, indirect, and informal analogies, one could argue that it remains unclear as to what the merit, game-theoretic or quantum physical, of such work is. What is needed in this context is a formal study of quantum games rooted in the formalism of cooperative game theory. 

Another interesting situation can be observed during the early years of quantum game theory (2002 to be exact) when Piotrowski \cite{PIOTROWSKI2002208} proposed a quantum physical model for markets and economics which are viewed as games. His ideas appear to be inspired by Meyer's work. In fact, Eisert {\it et al.}'s paper does not even appear as a reference in this paper. In a later paper \cite{PIOTROWSKI2003516} however, Piotrowski states ``We are interested in {\it quantum description of a game} which consists of buying or selling of market good'' (the emphasis is our addition). Note that from the terminology used in both of Piotrowski's papers, it seems that the author wants to implement games quantum physically, even though his initial motivation comes from gaming the quantum! This goes to show that in the early years of quantum game theory, the motivation for merging aspects of quantum physics and game theory was certainly not clear cut. 

Finally, there are offenders in the quantum physics community who use the word ``game'' colloquially and not in a game-theoretically meaningful way. An example can be found in \cite{Mariantoni_etal2011}. Such vacuous usage of the word ``game'' further confuses and obfuscates serious studies in quantized games and gaming the quantum, cooperative or not.

Literature on quantum games is considerable. A good source of reference is the Google Scholar page on the subject \cite{GoogleScholar} which contains a wealth of information on past and recent publications in the area. The survey paper by Guo {\it et al.} \cite{GUO2008318} is an excellent precursor to our efforts here.  

\subsection{Quantum entanglement: Nash versus social equilibrium}\label{social}

Entanglement in a quantum physical system implies non-classical correlations between the observable of the system. While Eisert {\it et al.} showed that their quantum computational implementation of Prisoner's Dilemma produced non-classical correlations and resolved the dilemma (Nash equilibrium is also optimal), in \cite{Shimumura2004}, Shimumura {\it et al.} establish a stronger result that entanglement enabled correlations always resolve dilemmas in non-zero sum games, and that classical correlations do not necessarily do the same. Quantum entanglement is clearly a resource for quantum games. 

In this section, we offer here a new perspective on the role of quantum entanglement in quantum games. We consider quantum entanglement in the context of Debreu's \cite{Debreu1952} ``social equilibrium''. Whereas Nash equilibrium is the solution of a non-cooperative game in which each player's strategy set is independent of all other players' strategy sets, social equilibrium occurs in a generalization (not extension) of non-cooperative games where the players' strategy sets are not independent of each other. These generalized games are also are known as {\it abstract economies}. Take for instance the example of a supermarket shopper (this example is paraphrased from \cite{Dasgupta2015}) interested in buying the basket of goods that is best for her family. While in theory she can choose any basket she pleases, realistically, she must stay within her budget which is not independent of the actions of other players in the economy. For instance, her budget will depend on what her employer pays her in wages. Furthermore, given her budget, which baskets are affordable will depend on the price of the various commodities, which, in turn, are determined by supply and demand in the economy. 

In an abstract economy, a player is restricted to playing strategies from a subset of his strategy set, with this limitation being a function of the strategy choices of all other players. More formally, in an abstract economy with $n$ players, let $S_i$ be the $i^{\rm th}$ player's strategy set and let $s_{-i}$ represent the $(n-1)$-tuple of strategy choices of the other $(n-1)$ players. Then player $i$ is restricted to play {\it feasible} strategies only from some $\gamma_i(s_{-i}) \subseteq S_i$ where $\gamma_i$ is the ``restriction'' function. In social equilibrium, each player employs a feasible strategy, and no player can gain by unilaterally deviating to some other feasible strategy. Debreu gives a guarantee of the existence of a social equilibrium in \cite{Debreu1952} using an argument that also utilizes Kakutani's fixed point theorem. Note that an abstract economy may be viewed as a type of mediated communication where the communication occurs via interaction with the social environment.

Recall that the EWL quantization of Prisoner's Dilemma utilizes maximally entangled qubits. Rather than as a quantum mechanism for mediated communication, we interpret the entanglement between qubits as a restriction function, restricting the players' strategy sets to feasible strategy subsets which Eisert {\it et al.} call the two-parameter strategy sets. It is exactly in this restricted strategy set that the existence of a dilemma-breaking, optimal Nash equilibrium is established. 

The point of note here is that the resource that quantum entanglement affords the players in the EWL quantization (and possibly others) can be interpreted in two different ways: one, as an extension to mediated quantum communication that produces {\color{magenta}a} near-optimal Nash equilibrium in the quantum game, and the other as a generalization of Prisoner's Dilemma to a {\it quantum abstract economy} with social equilibrium, where entanglement serves as the environment. In the former interpretation, Nash equilibrium is realized in mixed quantum strategies; in the latter interpretation, social equilibrium is realized via pure quantum strategies. Whereas the Nash equilibrium is guaranteed by Glicksberg's fixed point theorem, the question of a guarantee of social equilibrium in pure strategy quantum games is raised here for the first time. We conjecture that the answer would be in the affirmative, and that it will most likely be found using Nash embedding of $\mathbb{C}P^n$ into $\mathbb{R}^m$ similar to the one appearing in \cite{Khan2018}.

Finally, the interpretation of quantum entanglement as a restriction function also addresses van Enck {\it et al}'s criticism of the EWL quantization as blurring the distinction between cooperative and non-cooperative games.

\section{Quantum games, quantum algorithms, and protocols}\label{quantumalgo}

Aside from the seminal work of Meyer, research work into the potential use of quantum games for developing quantum algorithms and quantum communication protocols is  sparse in the early years of quantum game theory. However, as we indicated in the beginning of section \ref{sec::noncop}, the MW quantization protocol stands out in this context. Notwithstanding its lack of game-theoretic substance, this protocol can be interpreted as an instance of applying game-theoretic reasoning to a quantum physical protocol once the preferences of the players in Prisoner's Dilemma are applied to the elements of the computational basis of the Hilbert space, and the matter of a proper game extension is ignored. In this new light of gaming the quantum, the MW protocol becomes the first ever application of non-cooperative game theory for constructing a constrained optimal quantum computational protocol. The work of Piotrowski in \cite{Piotrowski2003} from 2003 discusses quantum games from a decision theory point of view, with examples from biology, economics, gambling, and has connections to quantum algorithms and protocols. A more recent work \cite{ZhangP} gives a secure, remote, two-party game that can be played using a quantum gambling machine. This quantum communication protocol has the property that one can make the game played on it demonstrably fair to both parties by modifying the Nash equilibrium point.

Some other recent efforts to game quantum computations as non-cooperative games at Nash equilibrium appear in \cite{FSKHAN2013, FSKHAN20131} and are reminiscent of work in concurrent reachability games \cite{BBMU-lmcs14, deAlfaro2007} and control theory in general. In \cite{Zabaleta}, the authors use quantum game models to design protocols that make classical communications more efficient, and in \cite{5453589} the authors view the BB84 quantum key distribution protocol as a three player static game in which the communicative parties form a cooperative coalition against an eavesdropper. 
In this game model, the coalition players aim to maximize the probability of detecting the intruder, while the intruder's objective is to minimize this probability. In other words, the BB84 protocol is modeled as a two player, zero-sum game, though it is not clear  whether the informal mingling of elements of cooperative and non-cooperative game theory skews the results of this paper. 

Further, in \cite{computation3040586}, the authors associate Meyer's quantum Penny Flip game with a quantum automaton and propose a game played on the evolution of the automaton. This approach studies quantum algorithms via quantum automaton akin to how behavioral game-theorists study repeated classical games using automaton such as Tit-for-Tat, Grimm, and Tat-for-Tit etc \cite{binmore2007playing}. Finally, some interesting studies into Meyer's Penny Flip game appear in \cite{Anand}, where the authors  consider a two player version of this game played with entangled strategies and show that a particular classical algorithm can beat the quantum algorithm, and in \cite{doi:10.1093/ptep/ptx182}, where the author formulates and analyzes generalizations of the quantum Penny Flip game to include non-commutative quantum strategies. The main results of this work is that there is sometimes a method for restoring the quantum state disturbed by another player, a result of strong relevance to the problem of quantum error-correction and fault-tolerance in quantum computing and quantum algorithm design. 

Finally, motivated by quantum game theory, in \cite{PhysRevA.95.062306} Bao {\it et al.} propose a quantum voting system that has an algorithmic flavor to it. Their quantized voting scheme enables a constitution to violate a quantum analog of Arrow's impossibility theorem, which states that every classical constitution endowed with three apparently innocuous properties is in fact a dictatorship. Thier voting system allows for tactical-voting strategies reliant on entanglement, interference, and superpositions and shows how an algorithmic approach to quantum games leads to strategic advantage.

Despite the excellent efforts of several authors in the preceding discussion, the approach of Meyer to search for quantum advantage as a Nash  equilibrium in quantum games remains largely unexplored. In terms of quantum communication protocols, where quantum processes are assumed to be noisy and are therefore modeled as density matrices or mixed quantum states, the Meyer-Glicksberg theorem offers a guarantee of Nash equilibrium. Whereas a similar guarantee in mixed classical states spurred massive research in classical computer science, economics and political science, the same is not true in the case of, in the least, quantum communication protocols. 

Likewise for pure quantum states. These states are used to model the pristine and noiseless world of quantum computations and quantum algorithms. One set of quantum computational protocols is the MW quantum game (not the MW game quantization protocol) which we interpreted above as having a Nash equilibrium in pure quantum states. Note that one could in principle say the same about the EWL quantum game, despite its questionable quantum physical reputation. However, unlike the situation with mixed quantum states, a guarantee of a Nash equilibrium has only come to light very recently \cite{Khan2018}. On the other hand, some efforts have been made in bringing together ideas from network theory, in particular network creation games \cite{Fabrikant:2003:NCG:872035.872088,Demaine:2012:PAN:2151171.2151176, Scarpa}, and quantum algorithms in the context of the quantum Internet and compilation of adiabatic quantum programs \cite{Khanieee}.

\section{Bell's inequalities and quantum games}\label{Bell}

It is well known that the EWL quantization scheme is limited in it's applicability to any situation where the players can perform any physically possible operation. This may be applicable in some instances when the hardware used to implement a quantum game only allows a limited set of operations, and there are no players with malicious intent or the technological sophistication to perform operations outside of the allowed set. However, for more sophisticated analyses that include such actors or other factors, a more general framework is desired. To this end, one can start by focusing on the role of quantum entanglement in quantum games. 

As discussed earlier, the most common interpretation is that the entanglement between players' quantum strategies acts as a type of mediated communication, advice, or contract between the players. A common objection is that quantum games have more strategy choices than the classical version, and that it is possible to simply reformulate a classical game to incorporate more strategy choices so that the classical game has the same equilibria as the quantum counterpart, as was shown with the quantum Prisoner's Dilemma \cite{PhysRevA.66.024306}. However, as is discussed below, this is not always the case. The study of Bayesian quantum games addresses these objections and has elucidated the role of entanglement in quantum games as well as the possible advantages of a quantum game by relating them to Bell's inequalities \cite{PhysRevA.96.042340}. 

The connection between Bell's inequalities and Bayesian quantum games was first recognized in the similarities between the form of a Bell's inequality and the payoff function of a Bayesian quantum game. It was found that by casting a quantum Bayesian game in a certain way, the payoff function can resemble a form of a Bell's inequality so that in the presence of quantum correlations, i.e. entanglement, the inequality will be broken and the quantum game will have a higher payoff than the classical version. In the analogy, a direct parallel can be drawn between the measurements and measurement outcomes in a Bell's inequality experiment to the player types and player actions in a Bayesian quantum game \cite{Brunner2013}. 

In Bayesian games, the players have incomplete information about the payoff matrix of the game. This can be formulated by assigning the players different {\sl types} characterized by different payoff matrices. When making their strategy choice, the players know their own type, but not that of their opponent \cite{Harsanyi1967}. This is also noted that this is related to the conditions in non-local games \cite{cleve2004consequences}, the condition that the players cannot communicate during a game, and the concept of the no-signaling in physics. A correspondence can be drawn between the condition of locality, used in deriving a Bell's inequality, and the condition that the players do not know the type of the other player. This condition can be described mathematically for a two player, two strategy game for example, by labeling the player types as $X,Y$ and the strategy choices as $x,y$ with the following equation\cite{Cheon2008}:

\begin{equation}
 P({x,y}|{X,Y}) = P(x|X)P(y|Y)
\end{equation}

That is, the probability of the joint actions ${x,y}$ given that the player types are $X$ and $Y$ is equal to the probability that a player of type $X$ plays $x$ multiplied by the probability that a player of type $Y$ plays $y$. The factorizability of the joint probability distribution is a statement that the players action cannot be influenced by the type of their opponent. It has been noted previously by Fine \cite{Fine1982} that a sufficient condition for the breaking of a Bell's inequality is that the joint probability distribution is non-factorizable. For example, if there are two players (X and Y), with two possible strategy choices ($x$ and $y$), the joint probability distribution of a mixed-strategy where both players choose each strategy with a  $50\%$ probability is given by:

\begin{center}
\begin{tabular}{ l || c | r }
$X| Y$ & $ (x)$ &$(y)$\\
\hline 
\hline
$(x)$& $ 0.25$ & $0.25$ \\
$(y)$& $0.25$ & $0.25$ 
\end{tabular}
\end{center}

If, for example, the players can base their strategy choice on the measurement of some quantum state, such as the entangled state $\frac{1}{\sqrt{2}}(\ket{x y} +\ket{yx})$, it is possible to realize the probability distribution:

\begin{center}
\begin{tabular}{ l || c | r }
$X| Y$ & $ (x)$ &$(y)$\\
\hline 
\hline
$(x)$& $ 0$ & $0.5$ \\
$(y)$& $0.5$ & $0$ 
\end{tabular}
\end{center}
This probability distribution, when analyzed for an individual player still has a $50\%$ probability of either strategy. The difference is that the strategy choices of X and Y are perfectly correlated. This probability distribution is not in the image of the original mixed strategy game and is not possible without some form of communication between the players or advice.
 
Thus it is possible to formulate a Bell's inequality from a given Bayesian quantum game and vice versa\cite{Silman2008,Flitney2009}. The objection that the strategy choices available to a quantum player are greater than that of the classical player was addressed by Iqbal and Abbott \cite{Iqbal2010}. They formulated a quantum Bayesian game using probability distributions rather than state manipulations. The condition of factorizability of these probability distributions produces constraints on the joint probability distributions of the players, which can in turn be formulated as Bell's inequalities. The advantage in this case is that the strategies available to the classical players are identical to those of the quantum players. The difference is that in the quantum case, the players are given an entangled input, while in the classical case they are given a product state. Within this formalism the solution to the Prisoner's Dilemma is identical in the quantum and classical case, whereas in other games, the violation of Bell's inequality can lead to a quantum advantage, as in the matching pennies game.

This analysis can be taken further to incorporate the player's receiving advice in a classical game. In a classical non-local game, the players are allowed to formulate strategies before the game and may be the recipients of some form of common advice, but the advice cannot depend on the state of nature. As discussed earlier, this leads to the correlated equilibria \cite{AUMANN197467}. As we also noted in section \ref{sec::noncop}, correlated equilibrium allows for the possible realization of more general probability distributions over the outcomes that may not be compatible with the player's mixed strategies. More precisely, a mixed strategy Nash equilibrium is a correlated equilibrium where the probability distribution is a product of two mixed-strategies. In quantum games, these non-factorizable probability distributions are provided by entanglement or mediated quantum communication.

Brunner and Linden\cite{Brunner2013} incorporated the correlations that can be produced from classical advice into their analysis of quantum games. In this case, the joint probability distribution can be described by: 

\begin{equation}
 P({x,y}|{X,Y}) = \int d \lambda \rho(\lambda) P(x|X,\lambda)P(y|Y,\lambda)
\label{eq:classical_advice}
\end{equation}

Where the variable $\lambda$ represents the advice, or information distributed ot the players according to the prior $\rho( \lambda)$. This type of probability model accurately describes the behavior of players under shared classical advice. This condition is precisely the condition that is used to derive a Bell's inequality, and the history of violation of Bell's inequalities shows that quantum correlations arising from entanglement can break the inequalities derived from equation \ref{eq:classical_advice}. Thus, entanglement produces joint probability distributions of outcomes that are not possible classically, not just because they are non-factorizable, but also because they cannot have arisen from a classical source of advice, or in traditional quantum mechanical terminology, a hidden variable. If these joint probability distributions are realized in a Bayesian game with payoffs assigned appropriately, the players with access to quantum resources can play with probability distributions that are more favorable than what is possible classically. 

Correlated equilibria are possible in classical games because of the existence of probability distributions that are not factorizable. They therefore exhibit a wider class of probability distributions. And indeed, Bell’s inequalities show that there are quantum correlation that are beyond the classically available correlations. Games designed around Bell’s inequalities demonstrate that there are quantum games that can out-perform even classical games with correlated equilibria. These games do not have the weakness of the EWL quantization schemes, where the same results can be obtained by allowing correlated equilibria and without restricting the allowed strategies, which can often make EWL games un-physical. More recently, several researchers have used these results to generate games based on Bell’s inequalities that exhibit true benefit from quantum correlations without suffering  from the shortcomings of earlier quantization schemes \cite{doi:10.1142/S0219749905000724, Brandenburger2016TeamDP, Zhang:2012:QSG:2090236.2090241, DBLP:journals/corr/AulettaFRSW16, PhysRevLett.114.020401}.

Thus it has been shown that the probability distributions of outcomes are more fundamental than the presence of entanglement within a game. Indeed these considerations shed light on other types of correlations that can exist, both within quantum mechanics and beyond quantum mechanics. For example, there are quantum states that exhibit quantum correlations even when the entanglement is known to be zero. These correlations are known as quantum discord, and it is possible to formulate games that have an advantage under quantum discord \cite{MeloLuna2016}. Further, there are types of correlations known to be consistent with the no-signaling condition that are not even possible with quantum mechanics, known as super-quantum correlations Popescu and Rohrlich \cite{Popescu1994}. Games formulated with these correlations can outperform even the quantum versions\cite{Brunner2013}.

The analysis of Bayesian quantum games has thus addressed several of the objections to the importance of quantum games. The correlations that exist, or the joint probability distributions, of mixed strategies are shown to be more powerful in analyzing the advantage of a quantum game than just the presence of entanglement. The connections between Bayesian quantum games and Bell's inequalities will likely continue to give insight to and play a role in analyzing either different games that are formulated, or forms of Bell's inequalities that are derived \cite{Guney2015}.

\section {Stochastic games}\label{Stochastic}

Stochastic games extend strategic games to dynamical situations where the actions of the players and history of states affect the evolution of the game. In this section let us review a subset of classical multi-stage stochastic games that are Markovian in evolution. The hope is that it will generate interest in quantizing such multi-stage games leveraging the advances in quantum stochastic calculus \cite{KP1992} and quantum stochastics \cite{HChang2015}. There is very little work done on quantum Markov decision processes (qMDP) \cite{Scott2014} which are specialized quantum games and so there are lot of opportunities to explore in this class of quantum stochastic games. We start our discussions with stochastic games, specialize them to Markov decision processes (MDP), review the literature on quantized MDPs that involve partial observations, introduce quantum probability and quantum Markov semigroups, and finally outline one possible way to quantize stochastic games. 

A classical stochastic game a la Shapely \cite{Shapley01101953} is a tuple $(\chi, A_i(x), Q_i(x, a), \mathscr{P}(x|x, a), \lambda, x^0)$, where $\chi$ is the finite state space, $A_i$ is the finite action space for individual players, $Q_i$ is the $i^{th}$ player's payoff function, $\mathscr{P}$ is the transition probability function which can be thought of as a random variable because it is a conditional probability and would become a completely positive map in the quantum case, $0 \leq \lambda \leq 1$ is the discount factor that is player $i$'s valuation of the game diminishes with time depending on this factor, and $x_0$ is the initial state of the game. The discount factor is introduced in infinite horizon games so as to have finite values and another way to understand it is to relate $\lambda$ to the player's patience. How much more does the player value a dollar today than a dollar received in the future which can be quantified by the factor, so as her discount factor increases she values the later amount more and more nearly as much as the earlier payment. A person is more patient the less she minds waiting for something valuable rather than receiving it immediately. In this interpretation higher discount factor implies higher levels of patience. There is yet another reason to discount the future in multi-stage games. The players may not be sure about how long the game will continue. Even in the absence of time preference per se, a player would prefer a dollar today rather than a promise of one tomorrow because of the uncertainty of the future. Put another way, a payoff at a future time is really a conditional payoff conditional on the game lasting that long.

The formulation of Shapely has been extended in different directions such as non-zero sum games, states that are infinite (countable as well as uncountable), and the existence of Nash equilibria established under some restricted conditions. For a recent perspective on dynamical games we refer the reader to Ref\cite{Eilon2015}.

The dynamic game \cite{Ramesh} starts at $x^0$ and all the players simultaneously choose apply a strategy $s_i$ that is an action from $A_i$ depending upon the history of states. The payoffs and the next state of the game are determined by the functions $Q$ and $P$. The expected payoff for player $i$ is given by
\begin {equation}
\Pi_i \left(s_1, s_2, \dots, s_n; x^0\right)= \mathscr{E}\left[\lambda^t\sum_{t = 0}^{\infty}{Q_i\left(s_1(x^t), s_2(x^t), \dots, s_n(x^t); x^t\right)}\right]. \label{Bellman}
\end {equation}
\begin {defn} A strategy is called a Markov strategy if $s_i$ is a strategy that depends only on the state and we will let $s_i(x)$ denote the action that player i would choose in state x.
\end {defn}
A Markov perfect equilibrium (MPE) is an equilibrium on the dynamic game where each player selects a strategy that depend only upon the current state of the game. MPEs are a subset of Nash equilibria for stochastic game. Let us start with the observation that if all players other than i are playing Markov strategies $s_{-i}$, then player i has a best response that is a Markov strategy. This is easy to see as if there exists a best response where player i plays $a_i$ after a history h leading to state x, and plays $a`_i$ after another history h` that also leads to state x, then both $a_i$ and $a`_i$ must yield the same expected payoff to player i. Let us define a quantity $V_i(x; s_{-i})$ for each state x that is the highest possible payoff player i can achieve starting from state x, given that all other players play the Markov strategies $s_{-i}$. A Markov best response is given by:
\begin {equation}
V_i(x; s_{-i}) = \max_{a_i \in {A_i(x)}} \mathscr{E}\left[ Q_i(a_i, s_{-i}(x); x) + \lambda\sum_{x`\in\chi}^{\infty}{\mathscr{P}(x`|a_i, s_{-i}(x), x)V_i(x`; s_{-i})}\right].
\end {equation}

\textbf {Existence of MPE for finite games} When the state space, number of players, and actions space are all finite a stochastic game will have a MPE. To see this, let us construct a new game with N*S players where N and S are the number of players and the states respectively of the original game. Then the payoff and action for player (i,x) are given by,
 \begin {align}
 a(i,x) &= a_i(x), \text{   } i \in {N}, x \in \chi.  \\
 R_{i,x} & = \mathscr{E}\left[\lambda^t\sum_{t = 0}^{\infty}{Q_i\left(a(1, x^t), a(2, x^t), \dots, a(n, x^t); x^t\right)}|x^0=x\right].
 \end {align}
 This is a finite stochastic game that is guaranteed to have a Nash equilibrium. It is also a MPE as each player's strategy depends only on the current state. By construction, the strategy of player i maximizes his payoff among all Markov strategies given $s_{-i}$. As  shown above each player i has a best response that is a Markov strategy when all opponents play Markov strategies.
 
\begin {defn} {Two player zero sum stochastic game:} A two player zero sum game is defined by an $m\times{n}$ matrix P, where the $P_{ij}$ corresponds to the payoff for player 1 when the two players apply strategies $i \in {A_1}, i=1,\dots{m}$ and $j \in {A_2}, j=1,\dots{n}$ respectively and correspondingly the payoff for the second player is $-P_{ij}$.
\end {defn}
When the players use mixed strategies $\Delta(S_1)$ and $\Delta(S_2)$ respectively, the game being finite {\color{magenta}is} guaranteed to have a Nash equilibrium as follows:
\begin {equation}
V(P) = max_{s_1 \in \Delta(S_1)} min_{s_2 \in \Delta(S_2)} s_1^\top{P}s_2 = min_{s_2 \in \Delta(S_2)} max_{s_2 \in \Delta(S_1)} s_1^\top{P}s_2.
\end {equation}
 The above mini-max theorem can be extended to stochastic games as shown by Shapley {\color{magenta}\cite{Shapley01101953}.}
 \begin {lem} Let A and B be two $m\times{n}$ matrices then $|val(B) - val(C)| \leq \max_{ij} |B_{ij} - C_{ij}|$ .
 \end {lem}
 Let $(s_1, s_2)$ and $(\hat{s_1}, \hat{s_2})$ be the NE for games B and C respectively. Then, it follows from the properties of NE that 
 \begin {align}
 s_1^\top{B}s_2 & \leq {s_1}^\top{B}\hat{s_2}. and \\ 
 s_1^\top{C}\hat{s_2} & \leq \hat{s_1}^\top{C}\hat{s_2}.  \\
 s_1^\top{B}s_2 + s_1^\top{C}\hat{s_2} & \leq {s_1}^\top{B}\hat{s_2} + \leq \hat{s_1}^\top{C}\hat{s_2}. \\
 s_1^\top{B}s_2 - \hat{s_1}^\top{C}\hat{s_2}  &\leq {s_1}^\top{B}\hat{s_2} - s_1^\top{C}\hat{s_2}. \\
 & \leq \max_{ij} |B_{ij} - C_{ij}|.
\end {align}
By a symmetric argument reversing B and C we establish the lemma. Let us now consider the stochastic version of this game played in k stages. The value of the game is defined via a function $\alpha_k:\chi\rightarrow{R}$ and operator $T$ which is a contraction as
\begin {align}
R_x(\alpha)(a_1, a_2) &= Q(a_1, a_2; x) + \lambda\sum_{x`\in\chi}\mathscr{P}(x`|a_1, a_2, x)\alpha_k, a_1\in{A_1(x)}, a_2\in{A_2(x)}. \\
\alpha_k(x) &= val(R_x(\alpha_{k-1}). \\
(T\alpha)(x) &= val(R_x(\alpha)).
\end {align}
To see that the operator $T$ is a contraction with respect to the supremum norm and thus the game has a fixed point with any initial condition let us consider
\begin {align}
\norm{T\alpha - T\alpha`}_\infty &= \max_{x\in\chi}|val(R_x(\alpha) - val(R_x(\alpha`))| \\
&\leq\lambda\max_{x\in\chi}\max_{a_1\in{A_1(x)}, x_2\in{A_2(x)}}|\sum_{x`\in\chi}\mathscr{P}(x`|a_1,a_2,x)(\alpha(x`) - \alpha`(x`))\\
& \leq \lambda \max_{x`\in\chi} |\alpha(x`) - \alpha`(x`)| \\
&= \lambda\norm{\alpha - \alpha`}_\infty
\end {align} 
\begin {thm} Given a two-player zero-sum stochastic game, define $\alpha^*$ as the unique solution to $\alpha^* = T\alpha^*$. A pair of strategies $(s_1,s_2)$ is a subgame perfect equilibrium if and only if after any history leading to the state x, the expected discounted payoff to player 1 is exactly $\alpha^*(x)$.

\proof. Let us suppose the game starts in state $x$ and player 1 plays an optimal strategy for $k$ stages with terminal payoffs $\alpha_0(x`) = 0, \forall{x`}\in\chi$ and plays any strategy afterwards. This will guarantee him this payoff
\begin {align}
val &= \alpha_k(x) - \frac{\lambda^k}{1 - \lambda}M. \\
M &= max_{x`\in\chi} max_{a_1\in{A_1(x`)},a_2\in{A_2(x`)}}|Q(a_1,a_2;x)|.
\end {align}
This follows from the observation that after $k$ stages the payoff for first player is negative of maximum possible for second player. When $k \rightarrow \infty$ the value becomes $\alpha^*$ and by symmetrical argument for layer two the theorem is established.
\end {thm}
\begin {prop}
Let $s_1(x), s_2(x)$ be optimal (possibly mixed) strategies for players 1 and 2 in zero- sum game defined by the matrix $R_x(\alpha^*)$. Then $s_1, s_2$ are optimal strategies in the stochastic game for both players; in particular, $(s_1, s_2)$ is an MPE.
\proof. Let us fix a strategy that could possibly be history dependent $\hat{s_2}$ for player 2. Then, we first consider a $k$ stage game, where terminal payoffs are given by $\alpha^*$. In this game, it follows that player 1 can guarantee a payoff of at least $\alpha^*(x)$ by playing the strategy $s_1$ given in the proposition, irrespective of the strategy of player 2. Thus we have:
\begin {equation}
\mathbb{E}\left[\sum_{t=}^{k-1}\lambda^t{Q}(s_1(x^t),\hat{s_2}(x^t)+\lambda^k\alpha^*(x^k)|x^0=x\right] \geq {\alpha^*}(x).
\end {equation}
From this we get
\begin {equation}
\mathbb{E}\left[\sum_{t=}^{k-1}\lambda^t{Q}(s_1(x^t),\hat{s_2}(x^t)|x^0=x\right] \geq {\alpha^*}(x) -\lambda^k\norm{\alpha^*}_\infty.
\end {equation}
we finally get an expression that in the limit $k\rightarrow\infty$ goes to $\alpha^*$ and by symmetric argument for the second player we establish the result. 
\begin {equation}
\Pi(s_1,\hat{s_2};x) \geq {\alpha^*}(x) -\lambda^k\norm{\alpha^*}_\infty - \frac{\lambda^k}{1-\lambda}M.
\end {equation}
\end {prop}

The method described above is the backward induction algorithm to solve the Bellman equation \eqref{Bellman} and is applicable for games with perfect state observation. For games with asymmetric information, that is{\color{magenta},} when players make independent noisy observations of the state and do not share all their information, we refer the reader to reference \cite{Nayyar2014} and the references mentioned therein. Our interest here is confined to games with symmetric information from the quantization point of view as they would be a good starting point.

Now, let us consider a special class of stochastic games called Markov decision processes (MDP) where only one player called MAX plays the game against nature, that introduces randomness, with a goal of maximizing a payoff function. It is easy to see that MDP generalizes to a stochastic game with two players MAX and MIN with a zero sum objective. Clearly, we can have results similar to stochastic games in the case of finitely many states and action space MDP with discounted payoff and infinite horizon.

\begin {prop} (Blackwell 1962\cite{Blackwell1962}) For every MDP $(\chi, A(x), Q(x, a), \mathscr{P}(x|x, a), \lambda, z)$ with finitely many states and actions and every discount factor $\lambda<{1}$ there is a pure stationary strategy $\sigma$ such that for every initial state $z$ and every strategy $\tau$ we have
\begin {equation}
v(xz, \lambda, \sigma) \geq {v(z, \lambda, \tau)}.
\end {equation}
Moreover, the stationary strategy $\sigma$ obeys, for every state z,
\begin {align*}
v(x^0, \lambda) &= (1 - \lambda)r(x^0,\sigma(z)) + \lambda\sum_{z`\in{\chi}}\mathscr{P}(z`|z,\sigma(z))v(z`,\lambda).\\
                          &= max_{a\in{A(z)}}\left((1 - \lambda)r(z,a)+\lambda\sum_{z`\in{\chi}}\mathscr{P}(z`|z,\sigma(z))v(z`,\lambda)\right).
\end {align*}
\end {prop}
 
\section {Quantum probability}\label{qprob}
Let us now review the basic concepts in quantum probability and quantum Markov semigroups that would be required to define quantum stochastic games. The central ideas of classical probability consist of random variables and measures that have quantum analogues in self adjoint operators and trace mappings. To get a feel for the theory, let us consider the most familiar example \cite{Obata2007} of random variables, namely the coin toss. 
\begin{defn} Pauli operators.
\begin{equation}
\sigma_x = \left( \begin{array}{cc} 0 & 1 \\ 1 & 0 \end{array} \right) \nonumber
\end{equation}
\begin{equation}
\sigma_y = \left( \begin{array}{cc} 0 & -i \\ i & 0 \end{array} \right) \nonumber
\end{equation}
\begin{equation}
\sigma_z = \left( \begin{array}{cc} 1 & 0 \\ 0 & -1 \end{array} \right) \nonumber
\end{equation}
\end{defn}
\begin{exmp} The probability measure of a random variable (r.v.) taking values +1 and -1 is defined by $\mu{=}(1/2)(\delta_{-1}+\delta_{+1})$.
The moment generating function of this r.v. is given by
\begin {align}\label{moment}
M^m(\mu) =  \int_{-\infty}^{+\infty}x^m\mu(dx) &= 1 \text{,               if m is even} \nonumber \\
         &= 0 \text{.               otherwise} 
\end {align}     
Let us now consider the self adjoint operator $A = \left( \begin{array}{cc} 0 & 1 \\ 1 & 0 \end{array} \right)$ and the canonical basis of $C^2, e_0=\left( \begin{array}{c}  1 \\ 0 \end{array} \right) \text{, }  e_1=\left( \begin{array}{c}  0 \\ 1 \end{array} \right)$. It is easy to show the following \\ 
\begin {align}\label{innerprod}
  \langle{e_0},A^m{e_0}\rangle &=   1 \text{,               if m is even} \nonumber \\
         &= 0 \text{.               otherwise} 
\end {align}    
From the equations (\ref{moment}) and (\ref{innerprod}), it is clear that the self adjoint Pauli operator $\sigma_x$ is stochastically equivalent to the Bernoulli random variable. This moment generating sequence can be visualized as a walk on a graph as follows:
\begin {equation}
A = A^+ {+} A^-{=} \left( \begin{array}{cc} 0 & 1 \\ 0 & 0 \end{array} \right) + \left( \begin{array}{cc} 0 & 0 \\ 1 & 0 \end{array} \right).
\end {equation}                   
Using this we can rewrite (\ref{innerprod}) as
\begin{equation}
 \langle{e_0},A^m{e_0}\rangle= \langle{e_0},(A^+{+}A^-)^m{e_0}\rangle=\sum\limits_{\epsilon_1,\dots,\epsilon_m\in\{\pm\}} \langle{e_0},A^{\epsilon_m}\dots{A^{\epsilon_1}}^m{e_0}\rangle.
\end{equation}     
\end{exmp}
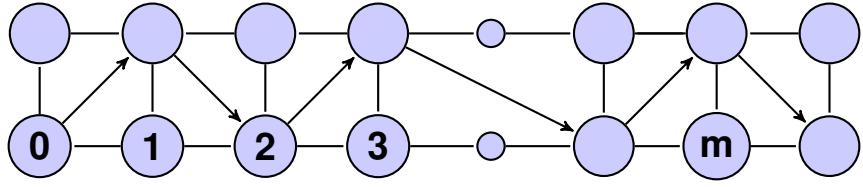
\begin{figure}
 \center {
 \begin{tikzpicture}[->,>=stealth',shorten >=1pt,auto,node distance=1.5cm,
  thick,main node/.style={circle,fill=blue!20,draw,font=\sffamily\Large\bfseries}]
  \node[main node] (1) {\hspace{0.5cm} };
  \node[main node] (2) [right of=1] {\hspace{0.5cm } };
  \node[main node] (3) [below of=1] {0};
  \node[main node] (4) [below of=2] {1};
  \node[main node] (5) [right of=2] {\hspace{0.5cm } };
  \node[main node] (6) [below of=5] {2};
  \node[main node] (7) [right of=5] {\hspace{0.5cm } };
  \node[main node] (8) [below of=7] {3};
  \node[main node] (9) [right of=7] {};
  \node[main node] (10) [below of=9] {};
  \node[main node] (11) [right of=9] {\hspace{0.5cm } };
  \node[main node] (12) [below of=11] {\hspace{0.5cm } };
  \node[main node] (13) [right of=11] {\hspace{0.5cm } };
  \node[main node] (14) [below of=13] {m};
  \node[main node] (15) [right of=13] {\hspace{0.5cm } };
  \node[main node] (16) [below of=15] {\hspace{0.5cm } };

  \path[every node/.style={font=\sffamily\small}]
    (1) edge[-] node {}  (2)    
    (2) edge[-] node {} (4)        
          edge[right] node[right=1mm] {}(6)
    (3) edge[-] node[right=1mm] {}(1)
          edge[left] node[right=1mm] {}(2)        
    (4) edge[-] node[right=1mm] {}(3)
         edge[-] node[right=1mm] {} (6)
    (5) edge[-] node[right=1mm] {} (6)
         edge[-] node[right=1mm] {} (2)
    (7) edge[-] node[right=1mm]{} (8)
         edge[-] node[right=1mm] {}(5)
    (6) edge[-] node[right=1mm] {}(8)
          edge[right] node[right=1mm] {}(7)
    (8) edge[-] node[right=1mm]{} (10)
    (7) edge[-] node[right=1mm]{} (9)
          edge[right] node[right=1mm]{} (12)
    (9) edge[-] node[right=1mm]{} (11)
    (11) edge[-] node[right=1mm]{} (12)
           edge[-] node[right=1mm]{} (13)
    (10) edge[-] node[right=1mm]{} (12)
    (12) edge[-] node[right=1mm]{} (14)
            edge[right] node[right=1mm]{} (13)
    (13) edge[-] node[right=1mm]{} (14)
            edge[right] node[right=1mm]{} (16)
            edge[-] node[right=1mm]{} (11)
    (15) edge[-] node[right=1mm]{} (16)
             edge[-] node[right=1mm]{} (13)
    (14) edge[-] node[right=1mm]{} (16);
    \end{tikzpicture}
  
  \caption{
     Coin-toss represented as a random walk on a graph with two nodes relating the random variable to the adjacency matrix of the graph.
  }
}
\end{figure}

\begin{defn} A finite dimensional quantum probability (QP) space is a tuple $(\mathscr{H},\mathbb{A}, \mathbb{P})$ where  $\mathscr{H}$ is a finite dimensional Hilbert space, $\mathbb{A}$  is a *-algebra of operators, and $\mathbb{P}$ is a trace class operator, specifically a density matrix, denoting the quantum state.\end{defn}
 
As we have alluded earlier, random variables in a CP are stochastically equivalent to observables in a Hilbert space $\mathscr{H}$. These are self-adjoint operators with a spectral resolution $X=\Sigma{_i}x_iE_i{^x}$ where the $x_i$'s are the eigenvalues of $X$ and each $E_i^X$ is interpreted as the event $X$ taking the value $x_i$. States are positive operators with unit trace and denoted by $\mathbb{P}$. In this framework, the expectation of an observable $X$ in the state $\mathbb{P}$ is defined using the trace as $tr \mathbb{P}(X)$. The observables when measured are equivalent to random variables on a probability space, and a collection of such classical spaces constitute a quantum probability space. If all the observables of interest commute with each other then the classical spaces can be composed to a product probability space, and the equivalence CP = QP holds. The main feature of a QP is the admission of possibly non-commuting projections and observables of the underlying Hilbert space within the same setting. 

\begin{defn} Canonical observables: Starting from a $\sigma$-finite measure space we can construct observables on a Hilbert space that are called canonical, as every observable can be shown to be unitarily equivalent to the direct sum of them \cite{KP1992}. Let $(\Omega, \Gamma, \mu)$, be a $\sigma$-finite measure space with a countably additive $\sigmaσ-algebra$. We can construct the complex Hilbert space as a space of all square integrable functions w.r.t $\mu$ and denote it as $L^2(\mu)$. Then, the observable $\xi^{\mu}:\Gamma\rightarrow\mathbb{P}(\mathscr{H})$ can be set up as  $(\xi^\mu(E)f)(\omega)=I_E(\omega)f[\omega],f\in{L^2}(\mu)$ where I is the indicator function. 
\end{defn}
\begin{exmp} Let  $\mathscr{H}$=$\mathscr{C}^2$ and $\mathbb{A}=M_2$ the *-algebra of complex matrices of dimension $2 \times 2$ and the state  $\mathbb{P}(A)=\langle\psi,A\psi\rangle=\langle{A^\dag}\psi,\psi\rangle$ where $\psi$ is any unit vector. This space models quantum spin systems in physics and qubits in quantum information processing. This example can be generalized to {\color{magenta} an} n-dimensional space to build quantum probability spaces. 
\end{exmp}
\begin{defn} Two quantum mechanical observables are said to be compatible, that is they can be measured simultaneously, if the operators representing them can be diagonalized concurrently. The two operators that share a common eigenvector will be characterized as co-measurable.
\end{defn}

There is a canonical way to create quantum probability spaces from their classical counterparts. The process involves creating a Hilbert space from the square integrable functions with respect to the classical probability measure. The *-algebra of interest is  usually defined in terms of creation, conservation, and annihilation operators. Classical probability measures become quantum states in a natural way through Gleason's theorem \cite{GL1957}. In this case, unitary operators are identified in the algebra to describe quantum evolutions. A sequence of operators forming a quantum stochastic process can be defined similarly to stochastic processes in a classical probability space.

Conditional expectation in a quantum context is not always defined and here we give a version that will be adequate for our purposes and is consistent with its classical counterpart in being a projection and enjoys properties such as tower. Followed by that we will define quantum Markov semigroups (QMS), as a one parameter family of completely positive maps, required for defining the quantum stochastic games. 

\begin {defn} A linear map $T:\mathscr{B}(\mathscr{H}_2)\rightarrow\mathscr{B}(\mathscr{H}_1)$ of bounded linear operators between Hilbert spaces $\mathscr{H}_2$ and $\mathscr{H}_1$ is called positive if
\begin {align*}
1. \quad T\mathbb{I} &= \mathbb{I}. \\
2. \quad TX &\geq {0} &\text{  whenever  } X \geq {0}. \\
3. \quad T(X^*) &= X^*. \\
4. \quad \norm{T(X)} &\leq \norm{X}. \\
5. \quad w.lim_{n\rightarrow\infty} T(X_n) &= T(X) &\text{  whenever  } w.lim_{n\rightarrow\infty} X_n = X.
\end {align*}
If $T^n, \forall{n}$ is positive then it is called a completely positive mapping.
\end {defn}

\begin{defn} A completely positive linear map from a *-algebra $\mathscr{A}$ to a von Neumann algebra $\mathscr{B}$ is said to be a version of quantum conditional expectation with respect to a state $\rho$ if it satisfies the following conditions:
\begin {align*}
\mathbb{E}_{\rho}[I|\mathscr{B}] &= \mathbb{I}.\\
\mathbb{E}_{\rho}[X^*X|\mathscr{B}] &\geq{0}, \forall{X}\in{\mathscr{A}}.\\
\mathbb{E}_{\rho}[X^*|\mathscr{B}] &= \mathbb{E}_{\rho}[X|\mathscr{B}]^*, \forall{X}\in{\mathscr{A}}.\\
\mathbb{E}_{\rho}[XYZ|\mathscr{B}] &= X\mathbb{E}_{\rho}[Y|\mathscr{B}]Z, \forall{X,Y}\in{\mathscr{A}}\text{ and } X,Y\in{\mathscr{B}}.\\
\mathbb{E}_{\rho}[\mathbb{E}_{\rho}[X|\mathscr{B}]] &= \mathbb{E}_{\rho}[X|\mathscr{B}], \forall{X}\in{\mathscr{A}}.
\end {align*}
\end{defn}

\begin {defn} A quantum Markov semigroup in a Hilbert space $\mathscr{H}$ is a one parameter family $\{T_t|t\geq{0}\}$ of operators, that is members of bounded linear operators $\mathscr{B}(\mathscr{H})$, satisfying the following conditions:
\begin {align*}
1. \quad T_0(X) =X, \forall{X}\in\mathscr{B}(\mathscr{H}), T_{t}T_s=T_{t+s} \text{  for all s,t  }\geq{0}. \\
2. \quad lim_{t\rightarrow{0}}\norm{T_t(X) - X} = 0, \forall{X}\in\mathscr{B}(\mathscr{H}). \\
3. \quad T_t(\mathbb{I}) = \mathbb{I}, \norm{T_t(X)}\leq\norm{X}, \\
 w.lim_{n\rightarrow\infty}T_t(X_n) = T_t(X) \hspace{2mm} {\rm \it whenever} \hspace{2mm} w.lim_{n\rightarrow\infty}X_n = X. \\
4. \quad{\rm for \hspace{1mm} any} \hspace{1mm} X_i, Y_i \in \mathscr{B}(\mathscr{H}), 1\leq{i}\leq{n}, n=1,2,\dots 
  \sum_{i,j}Y_i^*T_t(X^*_iX_j)Y_j \geq{0}.
\end {align*}
\end {defn}

Now, we have all the ingredients to define a multi stage qMDPs and quantum stochastic games. There could potentially be several ways to quantize a stochastic game and here we attempt at a straightforward generalization of the probability transition mapping to a QMS. We also introduce partial observation at every stage to evaluate the payoff. In this context it is interesting to note the ways in which random walks and classical filters are quantized. The quantum version of the former is without observations as that would reduce it to a classical walk and the later involves observing, more precisely non demolition measurements, compatible Hermitian operators to facilitate conditional expectations \cite{Luc2007}. Measurements may make quantum stochastic games behave similar to their classical counter part and coherent evolution of such games may result in interesting dynamics, for example, they may follow geodesics on the Bloch sphere not possible with measurement based controls \cite {Kurt2014}, and possibly with quantum advantage that require in depth investigations.
 
In this work {\color{magenta}we} consider games with long but finite horizon, that is the number of stages are finite and the payoffs are calculated at the last stage, and correspondingly the quantum measurements are made at the very end.

\section {Quantum Markov decision processes}\label{qMarkov}
A quantum Markov Decision Processes (qMDP) is a tuple $(\chi={\mathscr{C}^2}^{\otimes{n}}, A(x), Q(x, a), \mathscr{T}(x|\hat{x}, a), \\
\lambda, \rho_0)$. Here, $\chi$ is the finite $2^n$ dimensional complex Hilbert space, $A$ is the finite action space for the single player (unitary operators of the Hilbert space $\mathscr{C}^n$ such as the Pauli operators), $Q$ is the player's payoff function based on partial observation of the state, $\mathscr{T}$ is a completely positive mapping that would induce a quantum Markov semigroup when it is time dependent, $0 \leq \lambda \leq 1$ is the discount factor, and $\rho_0$ is the initial quantum state of the game. In terms of quantum information theory the state of the game can be represented by $n$ qubits and each player applies an unitary operator, a fixed finite set, on a qubit as a strategy.  

Instead of partially observed state based payoff, {\color{magenta}a} continuous non-demolition measurement based approach can be formulated. In \cite {Luc2007} Boutan et al., derived Bellman equations for optimal feedback control of qubit states using ideas from quantum filtering theory. The qubit is coupled to an environment, for example the second quantized electromagnetic radiation, and by continually measuring the field quadratures the state of the qubit (non-demolition measurements) can be estimated. The essential step involves deriving a quantum filtering equation rigorously based on quantum stochastic calculus \cite {KP1992} to estimate the state of the system coupled to the environment. By basing the payoff on this estimate the qMDP process can evolve coherently until a desired time. Such an approach can be extended to the quantum stochastic games described next. In addition, coherent evolutions may have advantages over measurement based dynamics \cite {Kurt2014}. 

\section {Quantum stochastic games}\label{QSG}
A quantum stochastic game is a tuple $(\chi={\mathscr{C}^2}^{\otimes{n}}, A_i(x), Q_i(x, a), \mathscr{T}(x|x, a), \lambda, \rho_0)$. Here, $\chi$ is the finite $2^n$ dimensional complex Hilbert space, $A_i$ is the finite action space for individual players (unitary operators of the Hilbert space $\mathscr{C}^n$ such as the Pauli operators), $Q_i$ is the $i^{th}$ player's payoff function, $\mathscr{T}$ is a completely positive mapping that would induce a quantum Markov semigroup when it is time dependent, $0 \leq \lambda \leq 1$ is the discount factor, and $\rho_0$ is the initial quantum state of the game. 

The quantum version of the dynamic game starts at $\rho_0$ which could be an entangled or a product state and all the players simultaneously choose to apply a strategy $s_i$, that is an action from $A_i\in{SU(2)}$, depending upon the history of states producing a quantum Markov semigroup. At each stage based on the strategies applied and the current state of the game, either by partial observation or via non demolition measurements on a common bath shared by the players, a referee moves it to the next step in a coherent fashion. The payoffs and the next state of the game are determined by the functions $Q$ and $\mathscr{T}$. The expected payoff for player $i$ given by
\begin {equation}
\Pi_i \left(s_1, s_2, \dots, s_n; \rho_0\right)= \mathscr{E}\left[\lambda^t\sum_{t = 0}^{\infty}{Q_i\left(s_1(x^t), s_2(x^t), \dots, s_n(x^t); x^t\right)}\right]. \label{QBellman}
\end {equation}

A quantum Markov perfect equilibrium (QMPE) is an equilibrium on the dynamic game where each player selects a strategy that depends only upon the current state of the game. QMPEs are again a subset of Nash equilibria for a quantum stochastic game. Let us reason as in the classical case that when all players other than player $i$ are playing Markov strategies $s_{-i}$, then player $i$ has a best response that is a Markov strategy. This follows as if there exists a best response where player $i$ plays $a_i$ after a history $h$ leading to state $x$, and plays $a`_i$ after another history $h`$ that also leads to state $x$, then both $a_i$ and $a`_i$ must yield the same expected payoff to player $i$. Let us define a quantity $V_i(x; s_{-i})$ for each state x that is the highest possible payoff player $i$ can achieve starting from state $x$, given that all other players play the Markov strategies $s_{-i}$. A Markov best response is given by:
\begin {equation}
V_i(x; s_{-i}) = \max_{a_i \in {A_i(x)}} \mathscr{E}\left[ Q_i(a_i, s_{-i}(x); x) + \lambda\sum_{x`\in\chi}^{\infty}{\mathscr{P}(x`|a_i, s_{-i}(x), x)V_i(x`; s_{-i})}\right].
\end {equation}
It is a straightforward procedure to start with a classical stochastic game first by constructing the canonical Hilbert space  \cite {HChang2015} with the probability measure $\mu$ and then generate the QMS as:
\begin {equation}
\mathscr{T}_t f(x) = \int_x f(y)Q_t(x, dy) = \int_x q_t(x, y)f(y)\mu(dy).
\end {equation}
With this set up we can leverage the mathematical machinery of QMS to investigate quantum stochastic games framed with different entanglement configurations, cost functions, etc,. Then, one can explore the condition under which Nash and Markov perfect equilibrium are possible for the quantum stochastic games. 

\section{Experimental realizations}\label{exp}

The implementation of quantum games on hardware can be viewed as a small quantum computation, in that sense, the requirements for a good platform on which to perform a quantum game are the same as that for a quantum computer\cite{DiVincenzio2000}.  The quantum computer may not need to be a universal computer but requires both single and two-qubit gates. Quantum computers with the capabilities required for quantum games are just beginning to come online (see for example www.research.ibm.com/ibm-q), and full quantum networks are in their infancy\cite{Pfaff2014}, thus, many of the experimental demonstrations to date, have been performed on hardware that are not ideal from the point of view of the criterion above.  Though notably, unlike many interesting quantum computing algorithms, quantum games are typically performed with very few qubits, making them an attractive application for early demonstrations on emerging quantum hardware.  

The potential applications and uses of quantum games suggest that certain characteristics are desirable for their implementation, beyond just those desirable for quantum computation. First, by definition, quantum games contain several independent agents. For realistic applications this likely requires that the agents be remotely located. This requires not only a small quantum computation, but also some form of network. The network would need to be able to transmit quantum resources, for example, produce entangled pairs that are shared at two remote locations which is typically done with photons. Second, a quantum game needs to have input from some independent agent, either a human or computer. This may require some wait time for the interaction with a classical system to occur, perhaps while the agent makes their decision. This implies that another desirable characteristic for quantum hardware is to have a quantum memory, that is, the ability to store the quantum information for some variable amount of time during the computation. Typically, the capabilities of a quantum information processor are quantified by the ratio of the coherence time to the time it takes to perform gates. Whereas in quantum games, the actual coherence time of the qubits may be a necessary metric in itself. 

One may wonder what it even means to perform an experimental demonstration of a quantum game. No experiment has ever had actual independent agents (i.e. humans or computers) play a game on quantum hardware in real time. Thus the implementations of games to date have been in some sense partial implementations. The games are typically implemented by using quantum hardware to run the circuit, or circuit equivalent, of the game with the strategy choices of the Nash equilibria. Where the strategy choices that form the Nash equilibria are determined by theoretical analysis of the quantum game in question. The output states of the experiment are then weighted by the payoff matrix, and the payoff at Nash equilibria is reported and compared to that of the classical case. One can view this as a type of bench-marking experiment where the hardware is bench-marked against a game theoretic application, rather than with random operations.
 
The games that have been implemented always set up to have a larger payoff in the quantum equilibrium than the classical case, presumably because these are the interesting games to quantum physicists. Because of this, the effect of noise or decoherence is almost always to lower the payoff of the players. It is generally seen that the payoffs at equilibrium of the quantum games still outperform the classical game with some amount of decoherence.

It should be noted that this section is concerned with evaluating the hardware for quantum games. As such the specific game theoretical results of the games that were performed will not be discussed, only the relative merits of each physical implementation.

\subsection{Nuclear magnetic resonance}

The first experimental realization of a quantum game was performed on a two qubit NMR quantum computer \cite{Du2002}. The computations are performed on the spin to spin interactions of hydrogen atoms embedded in a deuterated nucleic acid cytosine whose spins interact with a strength of $7.17 Hz$. They examined the payoff of the quantum game at Nash equilibrium as a function of the amount of entanglement. The experimentally determined payoffs showed good agreement with theory, with an error of 8 percent. In total, they computed 19 data points, which each took 300 ms to compute compared to the coherence time of the NMR qubits of $\sim 3$ seconds.

An NMR system has also demonstrated a three qubit game\cite{Mitra2007}. This game is performed on the hydrogen, fluorine and carbon atoms in a $^{13} CHFBr_2$ molecule. The single qubit resonances are in the hundreds of MHz, while the couplings between spins are tens to hundreds Hz. For their theoretical analysis they used three possible strategy choices, resulting in 27 possible strategy choice sets, which can be classified into 10 classes.  They show the output of the populations of all 8 possible states in the three qubit system, and thus the expected payoff, for each class of strategy choices sets. They ran the game 11 times, varying the amount of noise on the initial state, thus direcly showing the decrease of the payoff as the noise increases.  Their experimental populations had a discrepancy  of 15 to 20 percent with the theory.

Quantum computations on NMR based systems are performed on ensembles of qubits, and can have relatively large signal sizes. However, there do not appear to be promising avenues for scaling to larger numbers of qubits, or interfacing with a quantum communication scheme. Also, NMR computers are not capable of initializing in pure quantum states. Thus, methods have been developed to initialize the states to approximate states, \cite{Cory1998}, but there is uncertainty as to whether such mixed states actually exhibit entanglement or if they are separable \cite{Braunstein1999}. 

\subsection{Linear optics}
Implementing quantum games with optical circuits has several appealing characteristics. They do not suffer from the uncertainty in entanglement of NMR computing and can potentially have very high fidelities. Gates are implemented with standard optical elements such as beam splitters and waveplates. Also, since they are performed on photons, they can naturally be adapted to work with remote agents. 

One possible implementation is to use a single photon and utilize multiple degrees of freedom. Typically the polarization state of the photon is entangled with its spatial mode. In \cite{Zhang2008}, a heavily attenuated He-Ne laser was used as a single photon source. The single photon is input into a Mach-Zender interferometer where the two paths through the interferometer forms the first qubit pair, and the polarization state of the photon forms the second. They are entangled by splitting the photon into the two paths depending on its polarization. Gates are performed by single photon polarization rotations, i.e. adding waveplates to the photons path. They report an error in the experimentally determined payoff to the theoretical one of 1 to 2 percent. 


Rather than using the path of an interferometer as the spatial degree of freedom, one can also use the transverse modes of a beam \cite{Balthazar2015,Pinheiro2013b}. In these implementations, beams of light are incident on holographic masks to produce higher order transverse modes in a polarization dependent way.  These implementations have typically been done at higher light levels, i.e. $\sim mW$, and the beams are imaged on a camera to determine the steady state value of many photons being run in parallel.

Another possible implementation using linear optics utilizes cluster states. This has been done for a two player Prisoner's Dilemma game \cite{Prevedel2007}. The computation is performed with a four-qubit cluster state. Gates are performed by measurements of photons. Spontaneous parametric down conversion in a non-linear BBO crystal produces entangled photon pairs which are then interfered with beam splitters and waveplates. The creation of the four-qubit cluster state is post-selected by coincidence clicks on single photon detectors, so that runs are only counted if four single photo detectors registered a photon. The experimentalists can also characterize their output with full quantum tomography, and they report a fidelity of sixty two percent. 

Rather than producing cluster states, one can take the entangled photon pairs output from a non-linear crystal and perform gates in much the same way as they are performed in the single photon case\cite{Altepeter2009,Schmid2010}. These approaches have reported fidelities around seventy to eighty percent.  

A four player quantum game has been implemented with a spontaneous parametric down conversion process that produces four photons, in two entangled pairs {\color {magenta}\cite{Altepeter2009}}. Again the information is encoded into the polarization and spatial mode of the photons. This method, with two entangled pair inputs, can naturally be set up to input a continuous set of initial states. The initial entangled state in this implementation is a superposition of a GHZ state and products of Bell states. Again, the fidelities are reported to be near 75 percent which results in errors in the payoff at the equilibrium of about 10 percent.

Another example of a linear optical implementation sheds light on other types of correlations that can occur in quantum mechanics that are beyond entanglement, i.e. discord\cite{Zu2012}. To create states with discord, the measurements are taken with different Bell pairs, and then the data are partitioned out into different sets randomly, which produces a statistical mixture of entangled states. Such a mixture is known to have no entanglement as measured by the concurrence, but retains the quantum correlation of discord. The entangled pairs were produced from spontaneous parametric down conversion. This experiment reported a fidelity of 95 percent. Notably, even when there is no entanglement, the game can still exhibit a quantum advantage.

The linear optical implementations are promising because of their ability to perform games with remotely located agents. They are also capable of high fidelity quantum information processing. However they have drawbacks as well.  In order to run a different circuit, one must physically rearrange the linear optical elements such as waveplates and beamsplitters which could be done with liquid crystal displays or other photonic devices, though this could be difficult to scale up to implementations of more complicated games. In addition, the production of larger amounts of entangled photon pairs is experimentally challenging. These make it difficult to scale up implementations with linear optical circuits to more complicated games. In addition, purely photonic implementations have no memory, and thus may not be conducive to games that may require wait time for a decision to be made, or some sort of feed forward on measurements. 

\subsection{Other proposals}
There are many other potential platforms for quantum information processing and it is unclear which will be dominant in quantum computation. Ion trapped systems and cavity QED systems stand out as having all of the characteristics we desired in a quantum information processor specifically designed for implementing quantum games: they are potentially powerful quantum computers, they can have long memory times, and can be coupled to photonic modes for long distance communication. There have been proposals for implementations of quantum games on such systems\cite{Buluta2006,Shuai2009}.  

Ion trapped qubits can perform quantum computations with as many as five qubits with a very high fidelity \cite{Debnath2016}. In addition, ion trapped systems can also be coupled to single photons for entangling remote ions \cite{Hucul2015}. Cavity QED systems have a single atom, or ensemble of atoms, strongly coupled to a photonic mode. This allows the quantum information of the atomic system, which can be used for information processing, to be mapped to the photonic system for communication purposes with very high fidelity. Recently \cite{Solmeyer2018}, a Bayesian quantum game was demonstrated on a five-qubit quantum computer where the payoff, and phase-change-like behavior of the game were analyzed as a function of the amount of entanglement.

\subsection{Human implementations}
In addition to the bench-marking types of demonstrations described above, there is a separate interpretation about what it means to implement a quantum game, and that is having actual agents play a game with quantum rules. For real world applications of quantum games, it is interesting to speculate on whether or not it will be possible for players to effectively learn the correct strategies in a quantum game if they have no training in quantum theory. In fact, one of the biggest problems of classical game theory is that players do {\it not} act entirely rationally, and thus the equilibria of a game are only a guide to determine what real players will do. This problem may be exacerbated in quantum games by the fact that the players will likely have little or no knowledge of quantum mechanics or entanglement.  There have been a few experiments to research this question\cite{Chen2006,Chen2008}. 

Due to the limited availability of quantum hardware, in order to ease implementation, in these experiments the quantum circuits are simulated on a classical computer. Though a simulated quantum game will give the same outputs as a quantum computer, there are none of the benefits afforded by the absolute physical security of quantum communication protocols, which are likely a very desirable quality of quantum games. In addition, if the quantum games become sufficiently complex, it may not be possible to simulate them efficiently on a classical computer, as the number of states in a computation goes up as $2^n$ where $n$ is the number of qubits, as is well known in quantum computing. 

In \cite{Solmeyer2018}, the players were randomly paired and played the quantum Prisoner's Dilemma game, and the results were compared for classical versus quantum rules. They also performed one experiment where the players played repeatedly with the same partner. In the classical interpretation of the Prisoner's Dilemma, one can interpret the people who play the Pareto optimal strategy choice, even though it is not a Nash equilibrium, as altruistic. In any real instantiation of the game, there will be some players who play the altruistic option, even though, strictly, it lowers their individual payoff. As such, the prediction of the Nash equilibria from game theory can be interpreted as a guide to what players may do, especially in repeated games. 

When players played the game with quantum rules, the players tended to play the altruistic option more often than in the classical case, as is predicted by the Nash equilibria that occur in the quantum version of the game. This at least shows that players, who had no formal training in quantum mechanics, though had some instruction in the rules of the game, were capable of playing rationally, that is, maximizing their payoff. Interestingly, the game theory prediction for behavior was actually closer to the behavior of the players than when the behavior of players playing a classical game is compared to the classical theory. The players of the classical game played less rationally than those of the quantum game, and there was more variation between players in the classical versions. 

These results may suggest that the players have more preconceptions about the strategy choices in the classical version than in the quantum version, where the interpretation is more complicated. In the classical version, they can choose to cooperate or defect independently, while in the quantum version ultimately, whether or not they cooperate also depends on the strategy choice of their opponent. Preconceptions about the strategy choices in the classical game may provide influences beyond the desire to simply maximize ones own payoff and lead to larger deviations from the game theory prediction.

A full implementation of a quantum game with real players on quantum hardware has not been performed. Yet demonstrations of quantum game circuits on quantum hardware are compelling because they provide results that are interesting while only using small numbers of qubits. As quantum networks and quantum computers become more developed, we expect that quantum games will play a role in their adoption on a larger scale either as applications or as a diagnostic tool of the quantum hardware.

\section{Future applications for quantum game theory}\label{future}
Applications of conventional game theory have played an important role in many modern strategic decision-making processes including diplomacy, economics, national security, and business. These applications typically reduce a domain-specific problem into a game theoretic setting such that a domain-specific solution can be developed by studying the game-theoretic solution.  A well-known application example is the game of ``Chicken'' applied to studies of international politics during the Cold War \cite{SCHELLING1966}. In Chicken, two players independently select from strategies that either engage in conflict or avoid it. Schelling has cited the study of this game as influential in understanding the Cuban Missile crisis. More broadly, studying these games enables an understanding of how rational and irrational players select strategies, an insight that has played an important role in nuclear brinkmanship.
\par
The method of recognizing formal game-theoretic solutions within domain-specific applications may also extend to quantum game-theoretic concepts. This requires a model for the game that accounts for the inclusion of unique quantum resource including shared entangled states. For example, Zabaleta {\it et al.} \cite{Zabaleta2016} have investigated a quantum game model for the problem of spectrum sharing in wireless communication environments in which transmitters compete for access. Their application is cast as a version of the minority game put forward by Challet and Zhang \cite{Challet1997} and first studied in quantized form by Benjamin and Hayden \cite{PhysRevA.64.030301} and also by Flitney and Hollenberg \cite{Flitney2005}. For Zabaleta et al., a base station distributes an $n$-partite entangled quantum state among $n$ individual transmitters, i.e., players, who then apply local strategies to each part of the quantum state before measuring. Based on the observed, correlated outcomes, the players select whether to transmit (1) or wait (0). Zabaleta et al.~showed that using the quantum resource in this game reduces the probability of transmission collision by a factor of $n$ while retaining fairness in access management. 
\par
In a related application, Solmeyer et al.~investigated a quantum routing game for sending transmissions through a communication network \cite{Solmeyer2017}. The conventional routing game has been extensively studied as a representation of flow strategies in real-world network, for example, Braess' paradox that adding more routes does not always improve flow \cite{Roughgarden2006}. Solmeyer et al.~developed a quantized version of the routing game modified to include a distributed quantum state between players representing the nodes within the network. Each player is permitted to apply a local quantum strategy to their part of the state in the form of a unitary rotation before measuring. Solmeyer et al.~simulated the total cost of network flow in terms of overall latency and found that the minimal cost is realized when using a partially entangled state between nodes. Notably, their results has demonstrated Braess' paradox but only for the case of maximal and vanishing entanglement. If, and when, quantum networks become a reality, with multiple independent quantum agents operating distributed applications, quantum game theory may not only provide possible applications, but may also be necessary for their analysis.
\par
In the field of decision science, Hanauske et al.~have applied quantum game theory to the selection of open access publishing decisions in scientific literature \cite{Hanauske2007}. Motivated by the different publication patterns observed across scientific disciplines, they perform a comparative analysis of open-access choices using three different games: zero-sum, Prisoner's Dilemma and stag hunt. The formal solutions from each of these classical games provide Nash equilibria that either discourage open access publication or include this choice as a minority in a mixed strategy. By contrast, Hanauske et al.~found that quantized versions of these games which include distributed quantum resources yield Nash equilibria that favor open access publication. In this case, quantum game theory may provide a more general probability theory to form a descriptive analysis of a such socially constructed environments. In addition to decision making applications, game theory may also serve as a model for understanding competitive processes such as those found in ecological or social systems.
\par
 It is an outstanding question to assess whether quantum game theory can provide new methods to these studies. In addition to the study of classical processes, such as evolution and natural selection, quantum game theory also shows promise for the study of strictly quantum mechanical processes as well. In particular, several non-cooperative processes underlying existing approaches to the development of quantum technology including quantum control, quantum error correction, and fault-tolerant quantum operations. Each of these application areas require a solution to the competition between the user and the environment, which may be considered to be a `player' in the game theoretic setting. The solutions to these specific applications require a model of the quantum mechanical processes for dynamics and interactions which are better suited for quantum game theory. 
\par
A fundamental concern for any practical application of game theory is the efficiency of the implementation. A particular concern for {\color{magenta}a} quantum game solution is the relative cost of quantum resources, including entangled states and measurements operations. Currently, high-fidelity, addressable qubits are expensive to fabricate, store, operate, and measure, though these quantum resources are likely to reduce in relative cost over time. For some users, however, the expense of not finding the best solution will always outweigh the associated implementation cost, and the cost argument need not apply for those application where quantum game theory provides a truly unique advantage. van Enk and Pike have remarked that some quantum games can be reduced to similar classical game, often by incorporating a classical source of advice \cite{PhysRevA.66.024306}. The effect of this advice is to introduce correlations into the player strategies in a way that is similar to how a distributed quantum state provides means of generating correlated strategies. For example, Brunner and Liden have shown how non-local outcomes in Bell's test can be modeled by conventional Bayesian game theory \cite{Brunner2013}. This raises the question as to whether it is ever necessary to formulate a problem in a quantum game theoretic setting. As demonstrated above, there are many situations for which distributed quantum resources offer a more natural application, e.g., quantum networking, and such formulations are at least realistic if not necessary.  
\par
The current availability of prototype general-purpose, quantum processors provides opportunities for the continued study of quantum game theory. This will include experimental studies of how users interacts with quantum games as well as the translation of quantum strategies into real-world settings. However, quantum networks are likely to be needed for field testing quantum game applications, as most require the distribution of a quantum resource between multiple players. Alongside moderate duration quantum memories and high-fidelity entangling operations, these quantum networks must also provide players with synchronized classical control frameworks and infrastructure. These prototype quantum gaming networks may then evolve toward more robust routing methods.

\section{Conclusion}

We gave a review of theoretical and experimental developments in quantum game theory since its formal inception in 1999 to the current state of affairs. We reviewed future applications of quantum games with respect to technological and human implementations. In case of the former, quantum networks and quantum Internet-of-Things are of  immediate practical interest, where the distribution of a quantum resource among many players creates a competitive scenario. While full scale human implementation of quantum games has not been achieved, small scale experiments have been done and produce counter-intuitive outcomes such as a greater prevalence of altruism than in the classical game. A dicussion of quantum games and their role in the development of quantum algorithms and communication protocols was also presented, albeit concisely due to the limited work done in this area.

Classical stochastic games were also discussed. These games have also applications in networks, whenever there is a competition for resources, and the quantum counter part will have applications in networks \cite{deSousa2008} of quantum computing and sensing devices. We also reviewed the notion of quantum probability in the context of classical and quantum stochastic processes which model ``single'' player games in a dynamical setting such as Markovian processes. Quantum games in the setting of Bell's inequalities were also reviewed. 

We note that unlike other mergers of game theory and scientific disciplines like economics and evolutionary biology, quantum game theory has seen stagnated growth and limited success. Major reasons for this situation are the misconceptions and confusions outlined in sections \ref{sec::noncop} and \ref{coop} which appear to have discouraged physicists, particularly quantum information scientists, from pursuing the subject seriously. 

Finally, we showed how ideas from differential geometry and topology can be used to guarantee the existence of Nash equilibria in pure strategy quantum games using classic fixed-points such as the one's using in classical game theory and quantum games played with mixed strategies.

We conclude this essay with the hope that a new generation of researchers will find this discussion useful in contextualizing quantum game theory and find new and exciting applications of quantum games to physics.

\section{Acknowledgements}
 This material is based upon work supported by the U.S. Department of Energy, Office of Science Advanced Scientific Computing Research and Early Career Research programs.
  
 \bibliography{mybib}

\end{document}